\documentclass[twocolumn,showpacs,preprintnumbers,amsmath,amssymb,superscriptaddress,prb]{revtex4}
\usepackage{graphicx}
\usepackage{graphics}
\usepackage[dvips,usenames]{color}
\usepackage{dcolumn}
\usepackage{bm}
\usepackage{draftcopy}
\begin{document}
\title{The One-Particle Spectral Function and the Local Density of States in a Phenomenological Mixed-Phase Model for High-Temperature Superconductors}


\author{Matthias Mayr$^\star$}
\affiliation{Max-Planck-Institut f\"ur Festk\"orperforschung, 70569 Stuttgart, Germany.}

\author{Gonzalo Alvarez}
\affiliation{Computer Science \& Mathematics 
Division, Oak Ridge National Laboratory, \mbox{Oak Ridge, TN 37831}}

\author{Adriana Moreo}
\affiliation{Condensed Matter Sciences Division, Oak Ridge National Laboratory, Oak
Ridge, Tennessee 37831 and Department of Physics and Astronomy, The University of Tennessee, Knoxville,
Tennessee 37996}

\author{Elbio Dagotto}
\affiliation{Condensed Matter Sciences Division, Oak Ridge National Laboratory, Oak
Ridge, Tennessee 37831 and Department of Physics and Astronomy, The University of Tennessee, Knoxville,
Tennessee 37996}

\begin{abstract}

The dynamical properties of a recently introduced phenomenological model
for high temperature superconductors are investigated. In the clean limit,
it was observed that none of the homogeneous or striped states that are
induced by the model at low temperatures can reproduce the recent
angle-resolved photoemission results for LSCO (Yoshida {\it et al.},
Phys. Rev. Lett., {\bf 91}, 027001 (2003)), that show a signal with two branches in
the underdoped regime. On the other hand, upon including quenched disorder
in the model and breaking the homogeneous state into ``patches'' that are
locally either superconducting or antiferromagnetic, the two-branch
spectra can be reproduced. In this picture, the nodal regions are
caused by $d$-wave superconducting clusters. Studying the density of
states (DOS), a pseudogap is observed, caused by the mixture of the gapped
antiferromagnetic state and a $d$-wave superconductor. The local DOS
can be interpreted using a mixed-phase picture, similarly as observed in tunneling 
experiments. It is concluded that a simple phenomenological model for
cuprates can capture many of the features observed in the underdoped
regime of these materials.

    
\end{abstract}

%
\maketitle

\section{Introduction}
Since the discovery of the high-T$_c$ cuprates it has been widely suspected
that key information 
for the understanding of superconductivity lies in the curious regime located 
between the insulating (antiferromagnetic, AF) and metallic (superconducting, SC)
phases. 
Whereas these two phases are individually fairly well understood phenomenologically, 
the intermediate regime is,
even after several years of research, still to a large degree mysterious. 
Experiments have 
often only been able to decide what this state is not - for example, it is not 
a simple Fermi liquid or any other clearly defined, well-known state.
This has given rise to a variety
of proposals in describing this peculiar regime, often in terms of ``exotic'' order 
scenarios such as competing charge-density wave \cite{re:vojta02} (CDW) or pair-density wave states \cite{re:nayak00},
staggered-flux phases \cite{re:wen96}, spin-Peierls \cite{re:affleck87} states and, among others, 
orbital currents \cite{re:chakravarty01, re:varma99}. The much-debated pseudogap (PG) in the DOS 
that appears in this phase may then be regarded 
as a manifestation of a hidden order. Alternatively, it has been suggested 
that the strange phase interpolating between AF and SC states 
may be characterized by a particularly strong attraction 
between charge carriers, strong enough to drive $T_c$ down, and leading to 
a state of preformed, but yet uncondensed, pairs \cite{re:emery95}. 
In this scenario, 
the PG temperature $T_{\rm PG}$ signals the onset of pairing fluctuations. 
On the other hand, from the experimental viewpoint the regime between the AF and
SC phases is often described as ``glassy'', with slow dynamics \cite{re:hunt01}.
The long discussions on these issues show that
an understanding of the low hole-doped cuprates has not been
reached yet, and more work is needed.

In recent years, progress in both sample preparation and measurement techniques has led to 
very interesting experimental results for this fascinating phase. 
Especially important in this context are the insights gained 
via angular resolved photoemission spectroscopy (ARPES) \cite{re:damascelli03}, since 
this technique allows for a direct tracking of the Fermi surface (FS) - if it exists - 
and therefore provides crucial information for the evolution of the metallic phase 
from its insulating parent compound. 
Moreover, in the particular case of La$_{1-x}$Sr$_x$CuO$_4$ (LSCO), 
a relatively simple single-layer cuprate, those data have been obtained over the 
whole doping range $x$, starting from the antiferromagnetic Mott insulator 
at $x$=0 up to the ``optimal'' doping $x$=0.15 \cite{re:ino00}. 
These new results apparently show that portions of a FS (nodal regions)
are present even for
the smallest doping levels considered, such as $x$=0.03, where the material is $not$
superconducting \footnote{Very recent heat transport investigations of YBa$_2$Cu$_3$O$_{6.33}$ 
in the very underdoped non-SC regime have been interpreted as providing evidence of a state very similar to a 
$d$-wave SC, comparable as in LSCO (Mike Sutherland, S.Y. Li, D.G. Hawthorn, R.W. Hill, F. Ronning, M.A. Tanatar, 
J. Paglione, H. Zhang, Louis Taillefer, J. DeBenedictis, Ruixing Liang, D.A. Bonn, and W.N. Hardy, cond-mat/0501247 (2005)).}. 
This suggests that even the slightly doped Mott insulator is emphatically different from its half-filled parent compound,
and that it should be 
best understood as some sort of insulator with coexisting (embedded) metallic regions.
In fact, the doping evolution of ARPES data \cite{re:yoshida03} displays two
different branches, one evolving from the insulator deep in energy and the
second created by hole doping and containing nodal quasiparticles. 
In addition, it was found that the FS fragments, which are located around
$(\pi/2,\pi/2)$, do not 
expand as the {\it hole density} is increased - as one would expect for a conventional
metallic state - 
but rather acquire more spectral weight. Only if the {\it temperature} is increased
does this FS arc widen, reminiscent to the closing of the gap in superconductors.
As it will be argued in this paper, the most 
natural explanation for all these results is that there are already {\it
$d$-wave paired} quasiparticles present in the glassy phase. The
simplest such picture is one of
phase-separation (PS), where the metallic and insulating states inhabit spatially
separated (nanoscale) regions, a view recently introduced by Alvarez {\it et al.} \cite{re:alvarez04b}.
This explanation is against the exotic homogeneous
states proposed as precursors of the SC state, but nevertheless
the reader should note that a mixture AF+SC is highly nontrivial as well,
and in many respects it is as exotic
as those proposals. For instance, this state has ``giant'' effects, as
recently discussed \cite{re:alvarez04b}, results compatible with those found
experimentally by Bozovic {\it et al.} \cite{re:bozovic04} and others \cite{re:decca00, re:quintanilla03} in
the context of the giant proximity effect. 

One might also note that similar concepts have been proposed in the context of  
the metal-insulator (M-I) transition in the manganites, and have found widespread
acceptance there \cite{re:dagotto01, re:dagotto02}. 
For this reason, there exists a concrete possibility to frame the M-I 
transition in strongly correlated systems in a 
unifying picture of phase-separation and subsequent percolation. Indeed, such a
proposal has been made \cite{re:burgy01} (by some of the authors), 
claiming that such a mixed-phase state is the
consequence of impurity effects acting in regions of the phase diagram, 
where the AF and SC phase are very close in energy. 
In fact, quenched disorder has such an extraordinary and unusual influence
because of the possible first-order nature of the clean limit M-I transition (which is not 
experimentally accessible), 
and the effective low-dimensionality of the transition-metal oxides under consideration. 
Note that in some scenarios \cite{re:zhang97, re:chen03} a similar sharp transition AF-SC is invoked,
but the influence of quenched disorder has not been explored. In our case, quenched
disorder is {\it crucial} to produce the intermediate state between the AF and SC phases.

An important experimental tool in detecting coexisting phases is 
scanning tunneling microscopy (STM), which provided clear indication 
for mixed-phase states in 
manganites years ago \cite{re:dagotto01, re:dagotto02}. 
More recently, similar results have also been 
available in slightly underdoped Bi$_2$Sr$_2$CaCu$_2$O$_{8+\delta}$ (BSCCO) \cite{re:lang02}, 
suggesting the existence of metallic and insulating 
phases on nanoscale (typically 2-3 nm) regions. Unfortunately, it is not possible to extend 
these measurements into the strongly underdoped regime, as BSCCO becomes chemically unstable.
However, such STM results were recently reported by Kohsaka {\it et al.}\cite{Kohsaka04} 
for Ca$_{2-x}$Na$_x$CuO$_2$Cl$_2$ in the neighborhood of the 
superconductor-insulator transition point at $x$=0.08. 
The compound Ca$_{2-x}$Na$_x$CuO$_2$Cl$_2$ is 
structurally closely related to LSCO, and therefore offers a unique 
opportunity to compare 
real- and momentum-space imaging. Kohsaka {\it et al.} 
identified separate metallic (SC) 
and insulating areas of approximately 2${\rm nm}$ in diameter and demonstrated 
that the two phases in question are only distinguished by a different ratio 
of metallic and insulating clusters. Together with the BSCCO data 
{\it this points toward a mixed-phase description of high temperature superconductors}.

It is clear that both STM and ARPES method can be challenged since
they are surface probes and not bulk investigations. However, 
recent bulk measurements using Raman scattering have provided results that 
also may best be interpreted in terms of mixed-phase states and have given further support 
for such scenarios. Machtoub {\it et al.} \cite{re:machtoub05} have studied LSCO at $x$=0.12 in the
SC phase and in the presence of a magnetic field. The results were
interpreted in terms of an electronically inhomogeneous state in which the field
enhances the volume fraction of a phase with local AF order at the expense
of the SC phase. The same conclusions were reached upon studying neutron sacttering in underdoped ($x$=0.10) LSCO, 
where it was claimed that the observed appearance of an AF phase as the SC is suppressed by an applied magnetic field 
points towards coexistence of those two phases \cite{re:lake02}. Recent infrared experiments on the Josephson plasma resonance
in LSCO have also suggested a spatially inhomogeneous SC state \cite{re:dordevic03}. 
The same conclusion was reached investigating a high field magnetoresistance \cite{re:komiya04}. 
Studies by Keren {\it et al.} \cite{re:keren02} using 
$\mu$SR techniques have also led to microscopic phase separation, involving 
hole-rich and hole-poor regions.

Clearly, the notion of electronic phase separation 
\cite{re:emery90} has a long history in the cuprate literature, 
going back at least to the original proposal of the stripe state \cite{re:zaanen89, re:poilblanc89, re:emery93, re:emery94}. 
Although stripes 
as originally envisioned should only appear for very specific doping 
levels, this may simply be a 
consequence of the approximation used and one could expect stripe-like 
correlations or corresponding 
charge fluctuations to appear for a broad array of doping fractions. 
Although stripes are only one possible 
avenue to introduce mixed-phase tendencies, they have nevertheless attracted 
considerable interest. This is understandable since
in the mid 90's experimental evidence from inelastic neutron scattering 
seemed to demonstrate stripe-like charge-order in Nd-doped LSCO at hole doping $x$=1/8 \cite{re:tranquada95}. 
However, the more recent investigations mentioned here suggest a broader picture of phase
separation, with nanocluster ``patches'' of random sizes and scales,
rather than more organized quasi-one-dimensional stripes.
    
Even if a (general) mixed-phase scenario is 
accepted owing to experimental evidence, it is still not clear whether or not the 
electronic inhomogeneity is caused by the (supposed) inherent tendency of 
strongly-coupled fermionic systems 
to phase-separate or by the imperfect screening of the dopant ions 
due to the proximity of the Mott insulating phase and concomitant strong disorder potentials.  
Those two scenarios are not mutually exclusive, but put different emphasis on the aspect 
of random (chemical) disorder: whereas in the latter scenario it is thought that disorder is strong 
enough to overcome the tendency of fermionic ensembles to form a homogeneous system,
particularly when phases compete, 
in the first one chemical disorder may act as a mere catalyst of already present tendencies, 
possibly leading to a pinning of stripes/charge-depleted regions. If, as one might assume, 
underdoped cuprates lie in between those extremas, some materials could be more 
influenced by disorder 
than others. Clearly, one is certainly dealing with highly complex systems when studying
lightly-doped transition metal oxides.

Our goal here is to investigate the spectroscopic properties of systems with competing 
AF and SC order, and to compare them to recent ARPES and STM experiments, in order to develop a 
coherent understanding of cuprates from the mixed-phase scenario point of view. The present
paper builds upon the recent publication by Alvarez {\it et al.} \cite{re:alvarez04b}, where the proposed state
was described in detail and ``colossal'' effects for cuprates were predicted. The approach we followed
assumes as an experimental fact that AF and SC phases compete, and addresses how the interpolation from
one phase to the other occurs with increasing hole doping, once sources of disorder are introduced. It is
concluded here that recent LSCO ARPES results can be neatly explained by the mixed-phase scenario.
The paper is organized as follows: In Section II the formal aspects of the problem are
described, including models and approximations. Section III addresses the ARPES data for
(i) a variety of uniform states (showing that none fits the experimental results) and (ii)
in the presence of quenched disorder that leads to the mixed-phase state. Results for the
latter are found to reproduce experiments fairly well. In Section IV, our results for
the local DOS are presented and discussed. Conclusions are provided in Section V.

\section{Theory and Methods}
A minimum model for describing the interplay between AF, SC (and possibly charge-order (CO))
can be found in the extended Hubbard model $H_{\rm UV}$, defined as 
\begin{eqnarray}\nonumber
H_{\rm UV} &=& -t\sum_{<{\bf ij}>,\sigma}(c^\dagger_{{\bf i}\sigma}c^{}_{{\bf j}\sigma}+H.c.) - \sum_{{\bf i}\sigma} \mu_{\bf i} n_{{\bf i}\sigma}\\
  &+& \sum_{\bf i} U_{\bf i} n_{\bf i\uparrow}n_{\bf i\downarrow} -  \sum_{\langle \bf ij \rangle} V_{\bf ij} n_{\bf i}n_{\bf j},
\label{eq:htuv}
\end{eqnarray}
where $c_{{\bf i}\sigma}$ are fermionic operators on a two-dimensional (2D) lattice with $N$=$L$$\times$$L$ sites. $t$ is the hopping between 
nearest-neighbor (n.n.) sites ${\bf i}$, ${\bf j}$, and serves as the energy unit. $U_{\bf i}$ is the usual Hubbard term and $V_{\bf ij}$$>$0 
describes an effective n.n. attraction. The local particle density $n_{\bf i}$=$\sum_{\sigma}$$c^\dagger_{{\bf i}\sigma}$$c^{}_{{\bf i}\sigma}$ 
is regulated by the chemical potential $\mu_{\bf i}$, 
which, like the interactions $U_{\bf i}$ and $V_{\bf ij}$, is allowed to vary spatially. To address the properties of this model, 
the four-fermion terms in $H_{\rm UV}$ are subjected to 
a Hartree-Fock decomposition,  
\begin{eqnarray}\nonumber
n_{\bf i\uparrow}n_{\bf i\downarrow} & \rightarrow & n_{\bf i\uparrow}\langle n_{\bf i\downarrow}\rangle + \langle n_{\bf i\uparrow}\rangle n_{\bf i\downarrow} - \langle n_{\bf i\uparrow}\rangle \langle n_{\bf i\downarrow} \rangle, \\
n_{\bf i} n_{\bf j} & \rightarrow & \Delta_{\bf ij} c^\dagger_{\bf j\downarrow}c^\dagger_{\bf i\uparrow}+\Delta^\star_{\bf ij} c^{}_{\bf i\uparrow}c^{}_{\bf j\downarrow}-
|\Delta_{\bf ij}|^2,
\label{eq:decomp}
\end{eqnarray}
where we have assumed that the predominant tendency in the $V_{\bf ij}$ term is towards superconductivity rather than particle-hole pairing, 
which would favor charge-ordering or PS. 
After such a transformation, one is left with two new (site dependent) order parameters (o.p.), $m_{\bf i}$ and $\Delta_{\bf ij}$:
\begin{eqnarray}\nonumber
m_{\bf i} & \equiv & \langle n_{\bf i\uparrow} \rangle - \langle n_{\bf i\downarrow} \rangle\\
\Delta_{\bf ij} & \equiv & \hspace*{0.3cm} \langle c_{\bf i\uparrow}c_{\bf j\downarrow}\rangle.
\label{eq:orderpara}
\end{eqnarray}
$m_{\bf i}$ is the local magnetization, and $\Delta_{\bf ij}$ is the SC o.p., 
defined on the link ${\bf ij}$. The Hamiltonian $H_{\rm UV}$$\equiv$$H_{\bf HF}$=$H_{\bf HF}'$+$H^{cl}_{\bf HF}$ 
now is quadratic in electron operators and written as
\begin{eqnarray}\nonumber
H_{\bf HF}&=&-t\sum_{<{\bf ij}>,\sigma}(c^\dagger_{{\bf i}\sigma}c^{}_{{\bf j}\sigma}
+H.c.) -\sum_{{\bf i}\sigma} \mu_{\bf i} n_{{\bf i}\sigma} \\ 
& - & 
\sum_{{\bf \langle ij \rangle}}(\Delta_{{\bf ij}}
c_{{\bf i}\uparrow}c_{{\bf j}\downarrow}+H.c.) + \sum_{\bf i} U_{\bf i} m_{\bf i} s_{\bf i}^z + \nonumber \\
& + & \sum_{{\bf \langle ij \rangle}}V_{\bf ij}|\Delta_{\bf ij}|^2 +\- 1/4 \sum_{\bf i}U_{\bf i} (n^2_{\bf i}-m^2_{\bf i}), 
\label{eq:hamfermi0}
\end{eqnarray}
after we have introduced the local spin operator $s_{\bf i}^z$=$n_{{\bf i}\uparrow}$-$n_{{\bf i}\downarrow}$. 
$H_{\bf HF}$ is effectively a single-particle Hamiltonian with 2$N$ basis states, which can be readily diagonalized 
using library subroutines in the general case. It needs to be stressed that the third line in Eq.(\ref{eq:hamfermi0}) 
($\equiv$$H^{cl}_{\bf HF}$) contains $c$-numbers only, and no operator terms, unlike the first two rows ($\equiv$$H_{\bf HF}'$). It is, 
however, the interplay between the second line in Eq.(\ref{eq:hamfermi0}), which tends to {\it increase} the o.p. amplitudes and the third one, which 
enforces an energy penalization for large values of $|\Delta_{\bf ij}|$, $m_{\bf i}$ that determines their actual values. We have resorted to two different 
approaches in studying $H_{\bf HF}$, a conventional mean-field method and a Monte Carlo (MC) technique, both of which we will describe in detail below.      

\subsection{Mean-Field Method}
The mean-field method relies on the self-consistency condition Eq.(\ref{eq:orderpara}) to determine the appropriate values of 
$m_{\bf i}$, $\Delta_{\bf i}$, and also $n_{{\bf i}\sigma}$. 
Diagonalization itself amounts to performing a slightly modified Bogoliubov-de Gennes (BdG) transformation \cite{re:atkinson03, re:ghosal02, re:ichioka99}, 
where the electron operators $c_{{\bf i}\sigma}$ are expressed in terms of new quasiparticles $\gamma_{n\sigma}$,
defined as:
\begin{eqnarray}
c_{\mathbf i\uparrow}&=&\sum_{n=1}^{n=N}\{ a_n(\mathbf i)\gamma_{n\uparrow}-
b^{*}_{n+N}(\mathbf i)\gamma_{n\downarrow}^\dagger\},\nonumber\\ 
c_{\mathbf i\downarrow}&=&\sum_{n=1}^{n=N}\{ b_n(\mathbf i) \gamma_{n\downarrow}+
a^{*}_{n+N}(\mathbf i)\gamma_{n\uparrow}^\dagger\}.
\label{eq:bogoliubov}
\end{eqnarray}
$a_{n}({\rm \mathbf i})$ and $b_{n}({\rm \mathbf i})$ in (\ref{eq:bogoliubov}) are complex numbers 
and are chosen so that a Hamiltonian that is diagonal in $\gamma_{n\sigma}$ emerges. In this approach, the o.p. amplitudes are 
determined via the self-consistency 
condition (\ref{eq:orderpara}), which allows to express both $m_{\bf i}$ and $\Delta_{\bf ij}$ in terms of the wave functions $a_n$(${\bf i}$), $b_n$(${\bf i}$):
\begin{eqnarray}\nonumber
\Delta_{\bf ij} &=&\sum^{N}_{n=1} a^{}_{n}({\bf i})a^*_{n+N}({\bf j})\langle \gamma_{n\uparrow}\gamma^{\dagger}_{n\uparrow}\rangle - b^*_{n+N}({\bf i})b_n({\bf j})\langle \gamma^{\dagger}_{n\downarrow}\gamma_{n\downarrow}\rangle,\\
n_{{\bf i}\uparrow}&=& \sum^N_{n=1}|a_n({\bf i})|^2\langle \gamma^{\dagger}_{n\uparrow}\gamma^{}_{n\uparrow}\rangle + |b^{}_{n+N}({\bf i})|^2\langle \gamma^{}_{n\downarrow}\gamma^{\dagger}_{n\downarrow}\rangle, \nonumber \\
n_{{\bf i}\downarrow}&=& \sum^N_{n=1}|b_n({\bf i})|^2\langle \gamma^{\dagger}_{n\downarrow}\gamma^{}_{n\uparrow}\rangle + |a^{}_{n+N}({\bf i})|^2\langle \gamma^{}_{n\downarrow}\gamma^{\dagger}_{n\downarrow}\rangle.
\label{eq:self}
\end{eqnarray} 
where $\langle \gamma^{\dagger}_{n\sigma}\gamma^{}_{n\sigma} \rangle$=$\{ 1+e^{\beta E_{n\sigma}}\}^{-1}$, 
i.e. the Fermi function $f$ with properties $f$($x$)=1-$f$(-$x$), $\beta$=1/$T$ the inverse temperature and $E_{n\sigma}$ 
are the eigenvalues of (\ref{eq:hamfermi0}). For the time being, however, we will work at $T$=0, which simplifies those 
relations considerably, and also guaranties better convergence of the self-consistent loop. 

The self-consistent set of Eqs. (\ref{eq:orderpara}), (\ref{eq:hamfermi0}), (\ref{eq:bogoliubov}), and (\ref{eq:self}) 
can be solved in an iterative procedure, provided a meaningful starting 
wavefunction $\{$$a_{n}({\bf i}$)$\}_0$ is chosen. From the knowledge of the resulting eigenfunctions, all relevant 
observables such as the single particle spectral function 
\begin{equation}
A({\bf k}, \omega) = -\frac{1}{\pi}Im G({\mathbf k}, \omega),
\end{equation} 
where $G({\mathbf k}, \omega)$ is the (retarded) Green's function  
and the local DOS 
\begin{equation} 
N({\bf i}, \omega)  = -\frac{1}{\pi}Im G({\mathbf i}, \omega)
\end{equation}
can be 
calculated in a straightforward fashion. Those two observables are the crucial quantities with regards to ARPES and STM experiments, respectively. 
Although this approach is easy to implement and well-established, we want to 
mention here that it has its pitfalls, which lie in the correct choice of $\{$$a_{n}({\bf i}$)$\}_0$. This is particularly 
problematic in the regime of small hole doping $\delta$=1-$\frac{1}{N}$$\sum_{\bf i}n_{\bf i}$, $\delta$$\ll$1, 
which is known to lead to ``stripe'' states for not too small values of $U$, provided a sufficiently correct selection of the seed functions is made. 
The exact nature of those stripe states, however, might depend on the linear lattice dimension $L$; also a possible intermediate state 
between the undoped AF insulator and the stripe state is beyond the grasp of BdG (in the clean limit, at least). Of course, it is this 
particular regime that is of interest in this work - and for cuprates in general - and therefore we have in addition resorted to a novel, alternative 
approach, which does not suffer from the disadvantages of the self-consistent technique, but provides an accurate, unbiased solution at any 
temperature. This is the MC technique described in the next subsection.

\subsection{Monte Carlo Procedure}

For this purpose, Hamiltonian (\ref{eq:hamfermi0}) is studied using a conceptually different approach, which stresses the importance of the 
$c$-number terms $\sum_{{\bf ij}}V_{\bf ij}|\Delta_{{\bf ij}}|^2$, $\sum_{\bf i}$$\frac{U_{\bf i}}{4}m^2_{\bf i}$, 
which only play a minor role in the self-consistent approximation. The MC technique also offers the additional advantage of treating $\Delta_{\bf ij}$ as a 
complex variable, and allows us to write $\Delta_{\bf ij}$=$|\Delta_{\bf ij}|$$e^{i\phi_{\bf ij}}$, where $\phi_{\bf ij}$ is the phase 
associated with the bond between sites ${\bf i}$, ${\bf j}$. We will change our notation from here on 
and write $\Delta_{\bf ij}$=$\Delta_{{\bf i}\alpha}$, where $\alpha$ is a unit vector along the $x$ or $y$ directions. In a similar 
fashion, we write $\phi^{\alpha}_{\bf i}$ instead of $\phi_{\bf ij}$ and also assume $V_{{\bf i}\alpha}$=$V_{\bf i}$. 
Therefore, we regard $\Delta_{{\bf i}\alpha}$ as a site variable and, thus, finally 
write $\Delta_{{\bf i}\alpha}$=$|\Delta_{\bf i}|$$e^{i\phi^\alpha_{\bf i}}$.

Within the MC method, the partition function $Z_{\bf HF}$ 
pertaining to $H_{\bf HF}$, 
given as 
\begin{eqnarray}\nonumber
Z_{\bf HF} = \prod^{N}_{\bf i=1} & & 
\int^1_{-1}  dm_{\bf i}e^{\beta\sum_{\bf i}\frac{U_{\bf i}}{4}m^2_{\bf i}}
\int^{\infty}_0 d|\Delta_{\bf i}| e^{-\beta \sum_{\bf i} V_{\bf i}|\Delta_{{\bf i}}|^2}
\times \nonumber \\ 
& & \int^{2\pi}_0  d\phi^x_{\bf i}d\phi^y_{\bf i}Z_c(\{ m_{\bf i} \}, \{|\Delta_{\bf i}|\}, \{\phi^{x,y}_{\bf i}\}),
\label{eq:part}
\end{eqnarray} 
is calculated via a canonical MC integration over the amplitudes $m_{\bf i}$, $|\Delta_{\bf i}|$ and the phases $\{ \phi^{x,y}_{\bf i} \}$. 
In Eq.(\ref{eq:part}), the $U_{\bf i}$$n_{\bf i}$-term from Eq.(\ref{eq:hamfermi0}) has been absorbed into the local chemical potential. 
The purely electronic partition function $Z_c$=Tr$\{ e^{-\beta H_{\bf HF}'} \}$ is obtained after diagonalizing $H_{\bf HF}'$ for a given, 
fixed set of those parameters and this diagonalization is responsible for limiting the lattice sizes. 
In fact, we chose another simplification and perform the ``magnetic'' integration over the signs only, rather than the amplitudes as 
well. This amounts to replacing the o.p. $m_{\bf i}$ by an Ising spin $S^z_{\bf i}$, which can take only the values $\pm$1. 
We will also replace $U_{\bf i}$$m_{\bf i}$ by a single term $J_{\bf i}$, without loss of generality (i.e. a certain value of $J_{\bf i}$ corresponds to 
a certain value of $U_{\bf i}$ and vice versa). Compared to the standard procedure described above, 
the MC technique has some distinct advantages, in describing both the AF as well as the SC degrees of freedom (d.o.f.), 
since its results do not depend on the initial configuration.   
On the other hand, the MC technique requires a very large number of diagonalizations,  
typically 10$^6$ rather than the $\sim$10$^2$ iterations necessary to achieve self-consistency in BdG methods. 
This renders calculations on system sizes common in BdG ($N$$\sim$1000) impossible \footnote{Polynomial expansion methods would allow to
apply the MC technique to larger systems. For further information, see, G. Alvarez, C. Sen, N. Furukawa, Y. Motome and E. Dagotto, cond-mat/0502461; to be published in Comp. Phys. Comm. (Elsevier).}.


The quantities of interest, such as the one-particle spectral function and the 
local DOS, can be evaluated in a straightforward 
fashion, either directly or via the Green's function, which can be 
easily derived from the MC process. Similar techniques have been pursued 
in the double-exchange model, which is relevant for the manganites, 
and an in-depth description can be found in Ref.\onlinecite{re:dagotto01}.
Further (technical) information with respect to the calculations presented here 
is provided in the appendix.

\section{Spectral Functions in the Presence of Competing States}
In this section, we will present the analysis of the one-particle spectral function
for several regimes of the phase diagram of $H_{\bf HF}$. For general doping and interaction 
values this can only be done with the MC routine described above.  

\subsection{Clean System}

\begin{figure}
\includegraphics[width=7cm,clip]{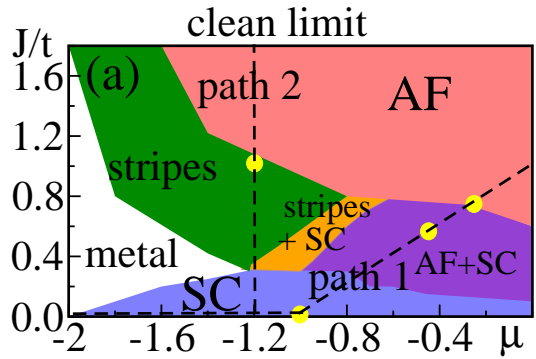}
\caption{\label{fig:cleanpd} MC phase diagram for Eq.(\ref{eq:hamfermi0}) 
without disorder at low temperatures, using 
$V$=1-$J/2$, reproduced from Ref.~\onlinecite{re:alvarez04b}. Five regions were observed: 
AF, $d$-SC, stripes, coexisting SC+AF, coexisting stripes+SC, 
and metallic. 
Yellow dots indicate where $A({\bf k},\omega)$ was calculated in the present work 
(see Fig.\ref{fig:akwcleanpath}).}
\end{figure}

The phase diagram of $H_{\bf HF}$ for the clean case (i.e. without quenched disorder) was presented in Ref.~\onlinecite{re:alvarez04b} and is
reproduced for the benefit of the reader in Fig.~\ref{fig:cleanpd}. 
The figure shows two ``paths'', which describe the transition from the AF to the SC phase.
The first one crosses a region of long-range order with \emph{local} AF/SC coexistence, whereas the second one 
involves an intermediate ``stripe'' state \cite{re:moraghebi02}. We do not discuss here the exact nature of the stripe state, which may be horizontal 
or diagonal, depending on parameters such as doping and lattice size. For our purposes it is sufficient that an inhomogeneous state 
- stripe, PS or CO - exists, and what its effects are with regards to experimental probes.  
Four representative points along those two paths (see Fig.\ref{fig:cleanpd}) were chosen and the corresponding 
spectral functions calculated.
 
\begin{figure}
\includegraphics[width=8cm,clip]{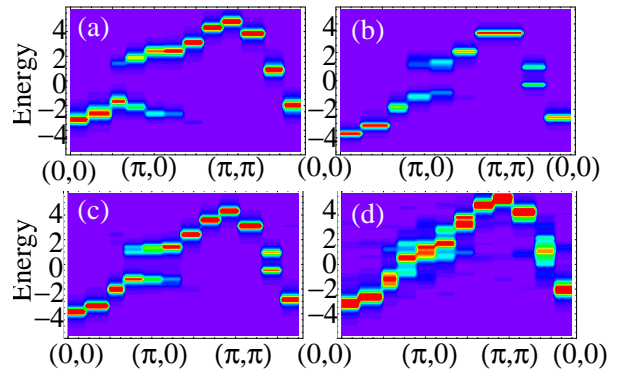}
\caption{\label{fig:akwcleanpath} $A({\bf k},\omega)$, evaluated via MC, on an 8$\times$8 lattice for  
(a) $(J,V,\mu)=(0,1,-1)$ (SC state), (b)  $(J,V,\mu)=(0.6,0.7,-0.4)$, 
coexisting AF/SC state, (c) $(J,V,\mu)=(0.7,0.65,-0.3)$ (AF), and (d) $(J,V,\mu)=(1,0.5,-1.2)$, striped state.}
\end{figure}

Figure~\ref{fig:akwcleanpath}(a) shows $A(\mathbf k,\omega)$ for the 
purely SC case ($J=0$) for $\mu=-1$, leading to a uniform density $\langle n\rangle\approx0.7$.
Fig.~\ref{fig:akwcleanpath}(b) is for the case when the system presents \emph{local} AF/SC coexistence (namely, both o.p.s 
simultaneously nonzero at the same site) and Fig.~\ref{fig:akwcleanpath}(c) 
for the pure AF phase. The red color indicates large spectral weight, whereas the blue one indicates very low intensity. 
In  Fig.~\ref{fig:akwcleanpath}(c), the AF gap can be clearly identified, together with the typical dispersion of the AF (upper branch), 
$E_{\bf k}$=$\pm$$\sqrt{\epsilon^2_{\bf k}+J^2}$, which makes $E_{\bf k}$ gapped everywhere. 
This is in stark contrast to Fig.~\ref{fig:akwcleanpath}(a), where
there are electronic states with appreciable intensity near the Fermi energy ($E_{\rm F}$) close to $(\pi/2,\pi/2)$, 
allowed by the symmetry of the pairing state. The ``intermediate'' state with local AF/SC coexistence is not drastically 
different from the one with AF correlations only, and its resulting 
energy dispersion can be simply described by $E_{\bf k}$=$\pm$$\sqrt{(e_{\bf k}-\mu)^2+J^2+\Delta^2_{\bf k}}$ once the parameter 
$\Delta_{\bf k}$ is known. This conclusion is not supposed to change using the SO(3)-symmetric spin model. 

Similarly, along path 2 of Fig.~\ref{fig:cleanpd} a point in the phase diagram with striped order was chosen, and the 
corresponding spectral density is given in  Fig.~\ref{fig:akwcleanpath}(d). This result compares very well with previous calculations, 
(Ref.~\onlinecite{re:moraghebi01}, Fig.~7): for instance, the system presents a FS crossing near $(\pi,0)$. Whereas the results from 
Fig.~\ref{fig:akwcleanpath}(a),(c) refer to generally well-understood phases of the cuprate phase diagram, Figs.~\ref{fig:akwcleanpath}(b),(d) 
are of relevance for the discussion related to the intermediate state, since they are both candidates for the intriguing phase in between.  
\begin{figure}[h]
\includegraphics[width=8cm,clip]{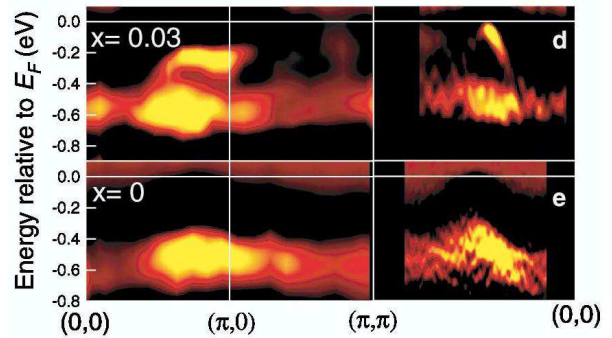}
\caption{\label{fig:yoshida03} Experimental ARPES spectra for LSCO with $x=0$ and $x=0.03$. Note the development of a (flat) second high-intensity branch 
near ($\pi$,0) and the emergence of a strongly dispersive signal at the Fermi level as the system is doped away from the half-filled insulator 
(reproduced from Ref.~\onlinecite{re:yoshida03}).}
\end{figure}

\begin{figure}
\includegraphics[width=8cm,clip]{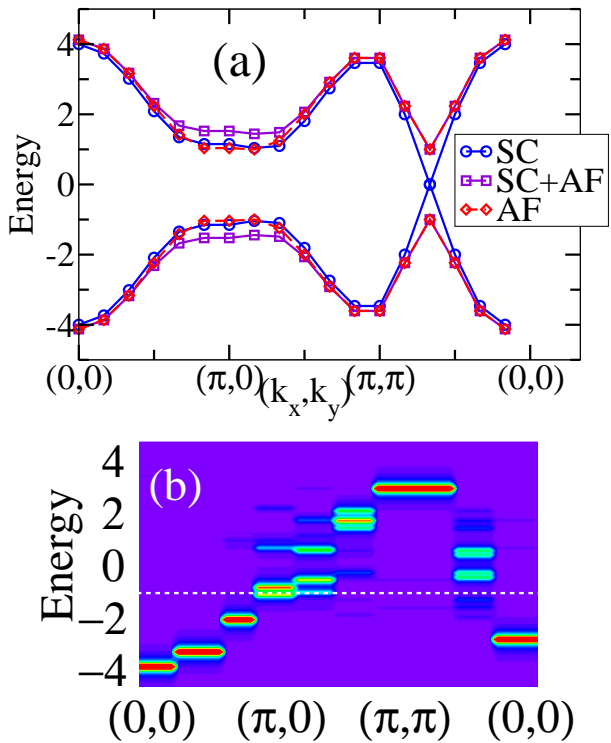}
\caption{\label{fig:akwcleanexact12x12} (a) Dispersion $E_{\mathbf k}=\pm\sqrt{(e_{\mathbf k}-\mu)^2+\Delta_{\mathbf k}^2+J^2}$
 for a perfect superconductor (SC) with $J=0$, $\Delta=0.3$,
 a perfect
coexisting superconductor and antiferromagnet (SC+AF) with $J=1$, $\Delta=0.3$, and a perfect AF with $J=1$, $\Delta=0$ as indicated.
$\Delta_{\mathbf k}=-2\Delta(\cos(k_x)-\cos(k_y))$ and $e_{\mathbf k}$ is the usual dispersion of the free system.
(b) Dispersion of a perfect striped configuration. Lattice sizes are 12$\times$12 and 8$\times$8, respectively.} 
\end{figure}

For comparison, ARPES data from Ref.~\onlinecite{re:yoshida03} for La$_{2-x}$Sr$_x$CuO$_4$ are reproduced in Fig.~\ref{fig:yoshida03}.
For very low doping $x=0.03$ (just inside the spin-glass insulating (SGI) phase) a {\it flat band} is observed close to -0.2eV in addition to a {\it lower branch}
(energy $\sim$-0.55eV), which is already present in the $x=0$ limit and therefore can be safely identified with the lower Hubbard band. 
As $x$ is increased even further, the lower branch retains its energy position, but gradually loses its intensity until it is almost completely invisible 
after the onset of the SC phase at $x$=0.06 \cite{re:ino00}. 
In contrast, the second branch gains in intensity with doping, and also moves continuously {\it closer} 
toward the Fermi level; at the same time it starts to develop a coherence peak, which is clearly visible for optimal doping.      
The main experimental result here, namely {\it the existence of two branches near $(\pi,0)$, cannot be reproduced using 
spatially homogeneous models} as demonstrated above. The cases of AF, SC and coexisting AF+SC states all show only one branch 
below $E_{\rm F}$ nearby $(\pi,0)$. This was already seen in Fig.~\ref{fig:akwcleanpath}(a)-(c) for the MC data and
is shown again in Fig.~\ref{fig:akwcleanexact12x12}(a) for all those configurations (in all these 
cases the exact dispersion is known).
 
If stripe configurations are considered, as in Fig.~\ref{fig:akwcleanpath}(d) (MC data) and Fig.~\ref{fig:akwcleanexact12x12}(b) 
(perfect configuration of stripes), there will appear two branches near $E_{\rm F}$, but the form of the dispersion is clearly different
from the experimental data in Fig.~\ref{fig:yoshida03}. The investigation of $A({\bf k},\omega)$ for a spin-fermion model, related 
to $H_{\bf HF}$, but retaining the SO(3) spin symmetry, has been done 
carefully in Ref.~\onlinecite{re:moraghebi01}. Again, stripe phases were found for certain parameters and while in some cases 
the existence of two branches near $(\pi,0)$ was reported, certainly there are no indications of ``nodal'' quasiparticles at $(\pi/2,\pi/2)$. 
Then, stripes alone are not an answer to interpret the results of Yoshida {\it et al}. As a consequence, we conclude that neither local AF+SC 
coexistence nor stripes can fully account for the ARPES results in the low-doping limit and alternative explanations should be considered. 

Beyond the results already described, ARPES also provides surprising insights/results for momenta other than ($\pi$,0) (Fig.\ref{fig:yoshida03}). 
Along the Brillouin zone diagonal, a dispersive band crossing $E_{\rm F}$ is found already in the SGI phase. 
The FS-like feature consists of a small arc centered at $\sim$ ($\pi$/2, $\pi$/2); surprisingly, 
as more holes are added, this arc does not expand, but simply gains spectral weight. This increase in spectral intensity is roughly 
proportional to the amount of hole-doping for $x$$\leq$0.1, although it grows more strongly thereafter. This observed increase in 
spectral weight is in relatively good agreement with the hole concentration $n_{\rm H}$ derived from Hall measurements and was interpreted as 
a confirmation of the hole transport picture. Below, however, we will provide a different explanation for this behavior.

The aforementioned large gap ($\Delta$$\approx$0.2eV) at ($\pi$, 0), 
together with the existence of the apparent gapless excitations around ($\pi$/2, $\pi$/2) is the essence 
of the PG problem. The shrinking of this gap and the concomitant appearance of a coherence peak 
has, for example, been interpreted as the evolution of a strongly coupled SC (at low doping) into a conventional BCS-SC at optimal doping. 
In this scenario, the large gap size directly reflects a large pairing scale, whereas the smallness of $T_c$ is attributed to the 
preponderance of phase-fluctuations in such a regime, which would outrule the existence of a phase-coherent SC 
condensate at higher temperatures. Alternatively, this gap may be regarded as the signal of a hidden order, which is not otherwise manifested. 
In other words, the relatively large excitation gap $\Delta_{\rm PG}$ is explained in terms of (i) a large SC 
gap $\Delta_{\rm PG}$=$\Delta_{\rm SC}$ itself, or (ii) $\Delta_{\rm PG}$=$\Delta_{\rm SC}$+$\Delta_{\rm ho}$, 
with a large, $x$-dependent hidden order gap $\Delta_{\rm ho}$   
whereas (iii) a mixed-state scenario, strongly influenced by disorder, leaves open the possibility 
that it is the (local) chemical potential that determines the PG physics. 
The precise role of $\mu$ in mixed-state phases needs to be examined further, but will not be addressed here.   

\subsection{Quenched Disorder I: Frozen Configurations}

\begin{figure}
\includegraphics[width=8cm,clip]{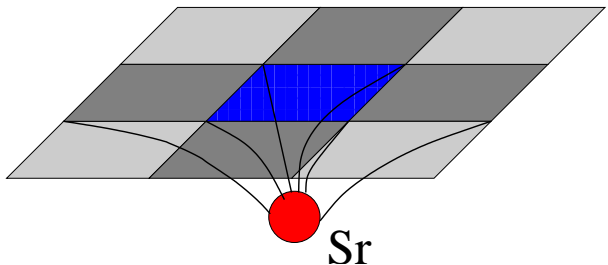}
\caption{\label{fig:plaquette}Schematic representation of Sr doping. A chemical dopant (Sr) will not only 
disorder the nearest sites (blue color) in the CuO$_2$-plane, but also neighboring ones, motivating the 
introduction of ``plaquette''-like disorder configurations.}
\end{figure}

\begin{figure}
\includegraphics[width=8cm,clip]{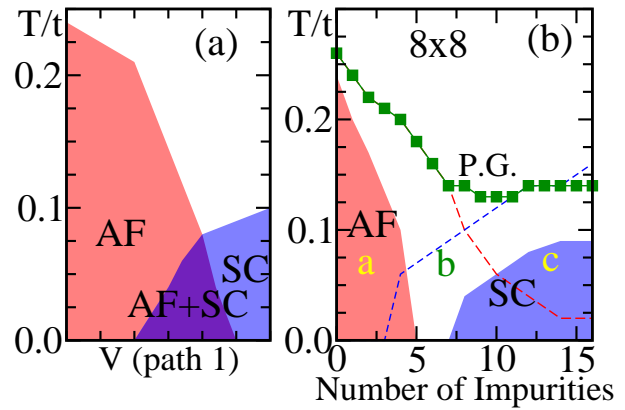}
\caption{\label{fig:pdafsc} Results reproduced from Ref.\onlinecite{re:alvarez04b} for the benefit of the reader. 
(a) Phase diagram of model Eq.~(\ref{eq:hamfermi0}) along Path 1 of Fig.~\ref{fig:cleanpd}, 
showing the AF, SC and local coexistence regions (for spatially constant couplings). (b) Same as (a) when quenched disorder is added to the
system in the form of SC plaquettes or impurities. In this case a region without long-range order appears. 
The temperature for PG formation is also indicated. Lattice size is 8$\times$8 in both cases. }
\end{figure}

Since calculations for   $A({\bf k},\omega)$ in the clean limit do not agree with 
ARPES measurements, we turn our attention to a system with quenched disorder. 
The impact of quenched disorder is realized by tuning the coupling constants $J_{\mathbf i}$ and $V_{\mathbf i}$ in Eq.(\ref{eq:hamfermi0}).
The spatial variations of these couplings is chosen in the following way (see Fig.\ref{fig:plaquette}): charge-depleted 
``plaquettes'' that favor superconductivity are placed on an AF background.
The same procedure was followed in Ref.~\onlinecite{re:alvarez04b}, where more details can be found. 
The main point is that impurities will not only influence a single site, but, possibly due to poor screening 
in the proximity of the insulating phase, will change local potentials over a rather large area. 
The phase diagram of the clean model along path 1 and the corresponding disordered case are reproduced in 
Fig.~\ref{fig:pdafsc} for the benefit of the reader. 
Disorder has opened a region between the SC and AF phases where none of the competing order dominates and 
both regimes coexist in a spatially separated, mixed-phase state. This ``glassy'' state was discussed in detail 
in Ref.~\onlinecite{re:alvarez04b}, where it was suggested that it leads to ``colossal effects'', in particular 
a giant proximity effect (GPE), which was recently observed in layered LSCO films \cite{re:bozovic04}. The pronounced susceptibility 
of such mixed-phase states towards applied ``small'' perturbations is well-known and is, e.g., often regarded 
as the driving force behind ``colossal magneto-resistance'' in manganites \cite{re:dagotto01}.

\begin{figure}
\includegraphics[width=8cm]{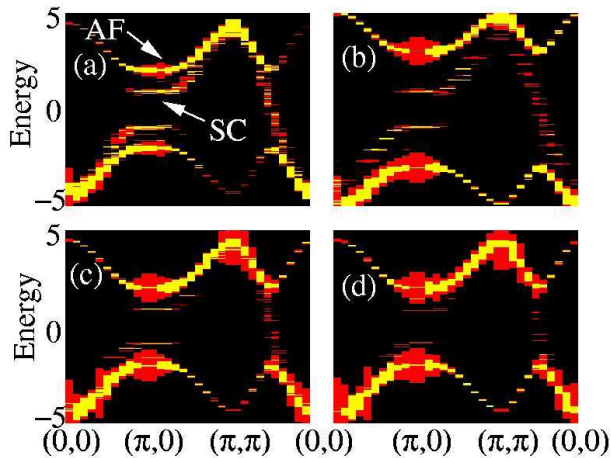}
\caption{\label{fig:l22x22arpes} Distribution of $A(\mathbf k,\omega)$ for a single configuration of classical fields, corresponding to a SC region
of size (a) 12$\times$12, 
(b)  8$\times$8, (c) 7$\times$7 or (d) 5$\times$5 on a 
22$\times$22 lattice (i.e., 30\%, 15\%, 10\% or 5\%  SC respectively). 
Shown is $E$ vs. ${\bf k}$ along $(0,0)\rightarrow(\pi,0)\rightarrow(\pi,\pi)\rightarrow(0,0)$.}
\end{figure}

To simplify the study and be able to access larger systems, 
we will first consider a single SC cluster embedded in an
AF background and also consider a fixed or ``frozen'' configuration of the classical fields (both AF and SC). 
Later, we will lift this restriction and perform a MC study. When a 12$\times$12 SC region is placed on an
AF background (total lattice size is 22$\times$22), the resulting distribution of $A(\mathbf k,\omega)$ is as shown in
Fig.~\ref{fig:l22x22arpes}(a). The contribution from the AF background is clearly distinguishable from that of the SC island, 
since it is present even when the SC region is removed. 
The SC cluster induces a second ``flat band'' - quite typical for gapped systems - near $E_{\rm F}$, along the $(0,0)\rightarrow(\pi,0)$ direction. 
That this flat band is indeed produced by the SC island is verified by decreasing the size of the island to
8$\times$8 (Fig.~\ref{fig:l22x22arpes}b),  7$\times$7 (Fig.~\ref{fig:l22x22arpes}c) and finally for 
 5$\times$5 (Fig.~\ref{fig:l22x22arpes}d), upon which this signal gradually decreases 
(the cases 9$\times$9 and 11$\times$11 give very similar results to 12$\times$12 and are
 not shown.). The spectral intensity related to the surrounding AF ``bath'' concurrently decreases, in agreement with experimental 
observations\cite{re:ino00}. 

\begin{figure}[h]
\includegraphics[width=6cm,clip]{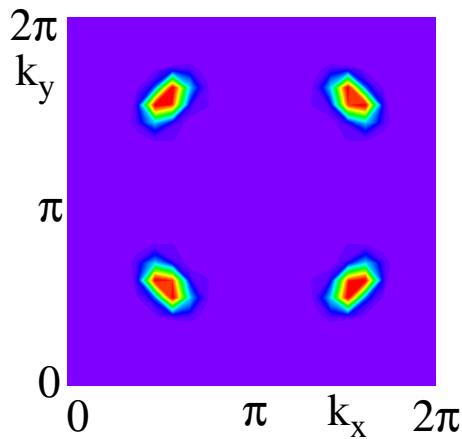}
\caption{\label{fig:l12x12arpeskxky} Energy cut of $A(\mathbf k,\omega)$ close to $E_{\rm F}$ for a single configuration of classical 
fields, corresponding to a SC region of size 12$\times$12 on a 22$\times$22 lattice. 
The brightness of the colors indicate the intensity in the $k_x$-$k_y$ plane for an energy $\omega=-0.24t$.} 
\end{figure}

The relative increase of the SC phase intensity also goes hand in hand with an increase in spectral intensity along 
the nodal direction, and the buildup of a FS around ($\pi/2$, $\pi$/2). Figure~\ref{fig:l12x12arpeskxky} shows a cut 
of $A(\mathbf k,\omega)$ near $E_{\rm F}$ for the case depicted in Fig.~\ref{fig:l22x22arpes}(a).  There is 
considerable spectral weight near the FS for momenta close to ($\pi$/2, $\pi$/2) only, suggesting that other parts of the FS are gapped. 

Therefore, even the simplest possible mixed-phase state can 
qualitatively account for the observed ARPES data. It is also interesting to note that SC signals comparably in strength with the ones stemming from 
the AF band, are only found for rather large SC blocks, encompassing at least 20$\%$ space of the whole system. From this point of view, even 
in the strongly underdoped limit at $x$=0.03, the relative amount of the SC phase has to be quite substantial already.  

\subsection{Quenched Disorder II: Monte Carlo Results}
The same system considered in the previous subsection was evolved by the MC procedure explained in the introduction.
However, the calculations can only be performed on smaller lattices. Results on a 10$\times$10
lattice were obtained using a 4$\times$4 region with couplings that favor superconductivity on a
background that favors antiferromagnetism. As observed in the previous subsection, the SC island produces the features seen near
$E_{\rm F}$, but on a 10$\times$10 lattice, and with a small 4$\times$4 SC island, 
there is not enough resolution to be able to see the ``flat band'' described previously. This makes it unlikely for such structures to be seen 
in ``exact'' solutions of Hubbard-type Hamiltonians, which, due to a variety of reasons, are restricted to lattices as large or smaller than the ones 
considered here. As a further consequence, it is inevitable to conclude that the SC islands in the real 
material must be substantially large, of sizes approximately $10\times10$ in lattice units which
translates into 40\AA$\times$40\AA\ nano clusters in physical units.

\subsection{Quenched Disorder III: Mean Field Theory}
The MC results presented above are also supported by more
traditional approaches, namely the solution of the BdG equations.
This self-consistent method allows for larger systems to be studied and is therefore much 
better suited to resolve the mostly subtle signals that are to be expected for the current investigation. 
In addition, the tracking of the small FS as the doping level is changed certainly requires a lattice grid that allows for a reasonable amount 
of resolution in k-space. This, of course, cannot, at the current time, be achieved with the MC technique presented above, yet 
this does not mean this approach is without merit for the problem at hand. 
One has to keep in mind that the MC approach can be justified on the grounds that it is unbiased towards any 
particular state - at least, if performed in a correct fashion - whereas the self-consistent solution 
may fail in this regard in the ``tricky'' regime of small, but finite, doping. The emergence of the stripe state in this case is well 
documented, but will only be realized for a sufficiently correct initial choice of the eigenfunctions.

With these caveats in mind, it may be the best strategy to employ MC to establish the correct (clean) phase diagram, and then using this information to 
describe the emerging phases with the help of BdG. In this spirit, we present some results for the inhomogeneous Hamiltonian $H_{\bf HF}$ below. 
Calculations were performed on systems up to $N$=32$\times$32. As before, plaquettes were inserted to create areas of predominant AF and SC 
order, respectively. These plaquettes were randomly distributed throughout the lattice, 
and were allowed to overlap. Owing to the random distribution of plaquettes, clusters of low 
electronic density are generated, their average size mainly determined 
by the the size of the underlying plaquettes (for the case of non-correlated plaquettes). 
We have chosen $U_{\bf i}$=5 (AF sites) and $V_{\bf i}$=-1 (SC sites), 
unless otherwise mentioned. With those values, it is possible to clearly separate the AF and SC signals, 
an important aspect when studying not overly large systems, and when one is looking 
for subtle signals. Our main conclusions, however, are not supposed to change for other parameter values. 
 
The number of plaquettes grows 
linearily with the hole density, which ranges from $x$=0.03 to $x$$\sim$0.2 to mimick ARPES investigations. The total percentage of the area occupied by 
SC bonds is roughly proportional to the hole density; however, given the phase diagram in Fig.\ref{fig:cleanpd}, there is a strong connection between 
the area, $a_{\rm SC}$, occupied by the SC regions, and the area occupied by the AF phase, $a_{\rm AF}$=1-$a_{\rm SC}$:
\begin{equation}
1-x=1 * a_{\rm AF} + n_{\rm SC}*a_{\rm SC},
\end{equation}  
where $n_{\rm SC}$ is the SC density (the AF density is n$_{\rm AF}$$\approx$1), which leads to the following expression
\begin{equation}
a_{\rm SC} = \frac{x}{1-n_{\rm SC}}.
\label{sc_ar_relation}
\end{equation} 
Eq. (\ref{sc_ar_relation}) can be interpreted in two ways: for a given value of $x$, either a desired value $a_{\rm SC}$ defines $n_{\rm SC}$, or vice versa. 
Here, our parameters (i.e. $\mu_{\bf i}$) are tuned so that $n_{\rm SC}$$\sim$0.75, the value upon where the system becomes 
a homogeneous SC according to Fig.\ref{fig:cleanpd}. 

\begin{figure}
\includegraphics[width=10.1cm]{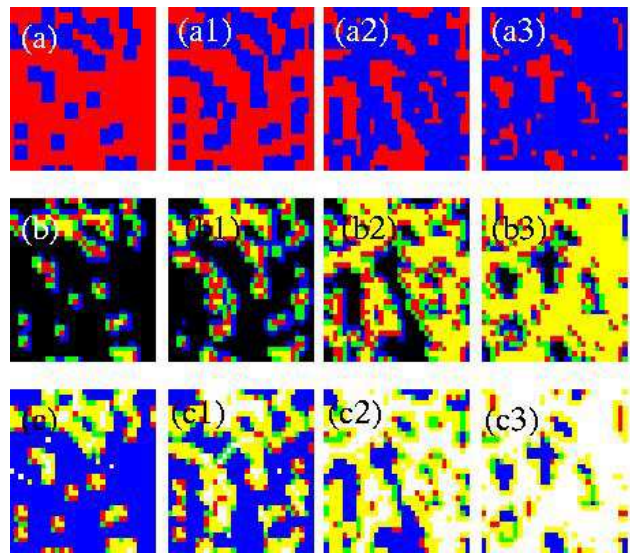}
\caption{\label{fig:definelabel} Typical quenched disorder configurations on a 32$\times$32 lattice 
at different electronic densities: the top row (a)-(a3) displays the local environments (red=AF, blue=SC), whereas (b)-(b3) shows the 
emerging local SC amplitudes, with the yellow color standing for the highest values (black for $\Delta_{\bf i}$=0). The bottom 
row shows the AF amplitudes. Blue equals strong AF ordering, whereas ever lighter colors identify a weak AF order parameter.}
\end{figure}

In Fig.\ref{fig:definelabel} we show typical configurations for which spectral functions were calculated, 
each corresponding to different densities, $\langle$n$\rangle$=0.97, 0.94, 0.87 and 0.79, respectively. 
The uppermost row ((a)-(a3)) displays the environment created by the plaquettes, with the red color 
favoring AF and the blue one the SC phase. In the second row ((b)-(b3)), the corresponding SC gap amplitudes are shown; yellow denotes 
large $\Delta_{\bf i}$'s (which are approximately in line with the ones found for a pure system at density $n_{\rm SC}$) and ever darker 
colors mean an ever smaller value of $\Delta_{\bf i}$. The bottom row plots the corresponding 
AF o.p., which, given our choice of plaquettes, results in what is essentially a mirror image of the second row. 
Here, blue denotes strong local AF amplitude $m_{\bf i}$, whereas lighter colors stand for very weak or absent magnetic ordering ((c)-(c3)). 
What the second and third row in Fig.\ref{fig:definelabel} show is that there are essentially three different regions in such a 
mixed-state: one that is a pure SC with no discernible AF amplitude, one where both amplitudes assume finite values (those may vary quite 
distinctively) and a third one that is solely AF. Those intermediate regions appear even though the underlying plaquettes are either 
purely AF or SC (which probably is an oversimplification to begin with), and owe their existence to boundary effects.

\begin{figure}
\includegraphics[width=9cm,clip]{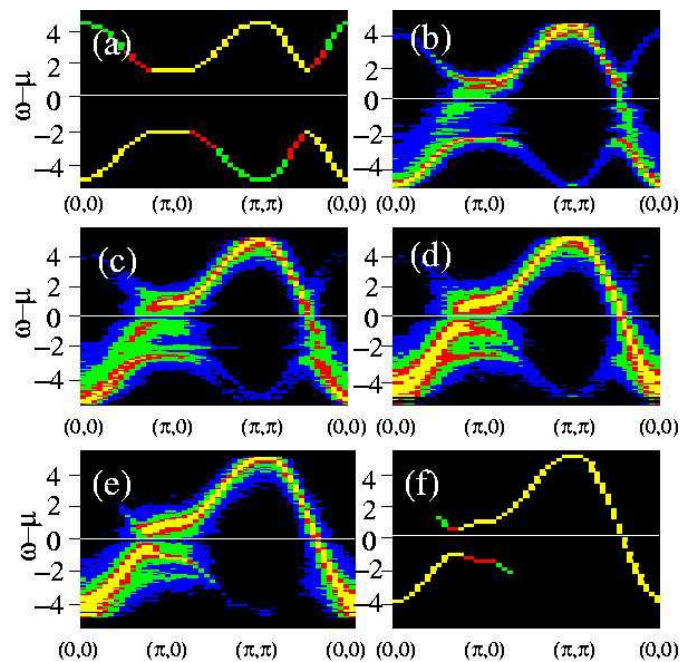}
\caption{\label{fig:arpes_bdg} $A(\mathbf k,\omega)$ vs. ${\mathbf k}$ calculated for a series of densities on a 32$\times$32 cluster, starting from 
$x$=0.0 (100$\%$ AF) in (a), to $x$=0.03 (22$\%$ SC) (b), $x$=0.06 (43$\%$ SC) (c), $x$=0.13 (70$\%$ SC)(d), $x$=0.21 (84$\%$ SC) (e), 
and the $x$=0.24 (100$\%$ SC) (f). The two branches visible near ($\pi$,0), belonging to the SC and AF region, 
respectively, can be clearly identified for the intermediate states, which suggests a simple mixed-state interpretation for ARPES experiments.}
\end{figure}

The spectral functions related to the mixed AF/SC states such as in Fig.\ref{fig:definelabel} are shown in Fig.\ref{fig:arpes_bdg}, in addition with 
those for the pure AF ($x$=0) case (Fig.\ref{fig:arpes_bdg}(a)) and the pure SC ($n_{\rm SC}$=0.76, Fig.\ref{fig:arpes_bdg}(f)); 
the former has the usual dispersion, characterized by a flat band in the neighborhood of the $X$ point, plus the shadow bands near ($\pi$, $\pi$). 
In Fig.\ref{fig:arpes_bdg}(a)-(f), the yellow color stands for large spectral weight 
(with respect to the maximum of $A({\mathbf k}, \omega)$ for each density), 
with ever darker colors describing ever smaller intensities. 
With the introduction of SC plaquettes to the AF background (Fig.\ref{fig:arpes_bdg}(a)$\rightarrow$(b)), spectral intensity is accumulating 
close to $E_{\rm F}$ at momenta ($\pi$, 0) and at $E_{\rm F}$ for $\sim$($\pi$/2, $\pi$/2), both of which 
were gapped at $x$=0. At the same time, the dispersion arising from the lower Hubbard band remains clearly identifiable 
and almost unchanged in position at energy $\omega_{\rm AF}$$\approx$-2$t$. 
Therefore, {\it in the neighborhood of the $X$ point two distinctive branches are emerging for 
phase-separated systems} such as in Fig.\ref{fig:definelabel} and both the SC and the AF branch can be clearly identified by 
comparison with the respective ``clean'' regime ($x$=0, $x$$\sim$0.2). 
Their relative intensities depend on the ratio $a_{\rm AF}$/$a_{\rm SC}$, with the AF 
branch becoming increasingly faint as this ratio tends towards 0. The observed features are very {\it broad} for 
weak doping for any wavevector close to $E_{\rm F}$ (Fig.\ref{fig:arpes_bdg}(b),(c)), characteristic of poorly defined quasiparticles with very small residue $Z$. 
This residue, however, appears to be small because of the disordered nature of the ground state, rather than 
as a consequence of strong interactions between the charge carriers - quite an important distinction! 
For example, as one moves closer towards the homogeneous state (Fig.\ref{fig:arpes_bdg}(c)$\rightarrow$(d)), 
the intensity distribution at ($\pi$, 0) becomes much sharper and reminiscent of a quasiparticle peak, 
without changing the interactions at all. 
All those observations are in agreement with Fig.\ref{fig:l22x22arpes} (although those data are much better defined, presumably 
because the o.p.'s were assumed constant for each region), as well as with experimental evidence, which has uncovered a 
very similar behavior in a series of underdoped to optimally doped LSCO \cite{re:ino00}.    

\begin{figure}
\includegraphics[width=8cm,clip]{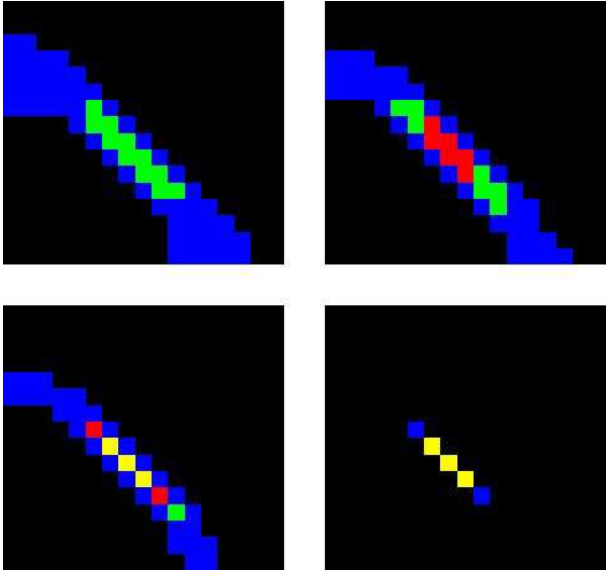}
\caption{\label{fig:fs_arc} The spectral intensities, integrated over a small shell below the Fermi level as described in the text, 
at different densities $\langle$$n$$\rangle$=0.97 (a), $\langle$$n$$\rangle$=0.90 (b), $\langle$$n$$\rangle$=0.87 (c) 
and $\langle$$n$$\rangle$=0.79 (d), and using a 32$\times$32 lattice. The FS arc developing around $\sim$ (0.45$\pi$, 0.45$\pi$) can be readily identified. 
It increases in intensity only as holes (plaquettes) are added to the system, whereas the arc's length remains essentially constant. 
This is to be compared with ARPES measurements in underdoped LSCO \cite{re:yoshida03}.}
\end{figure}

Similar agreement between theory and experiment is found around the ($\pi$/2, $\pi$/2) point, 
where significant spectral weight akin to a quasiparticle peak 
is found right at the Fermi level, and this weight continously increases as $x$ grows. 
This can be seen in Fig.\ref{fig:arpes_bdg}, but is more obvious in Fig.\ref{fig:fs_arc}, 
which shows the spectral intensities, integrated between the chemical potential and a 
cutoff $\omega_{\rm D}$=-0.3t, at different densities. Note that these data 
are calibrated with respect to the maximum intensity as found in Fig.\ref{fig:fs_arc}(d). 
The emergence of a FS 'arc', centered in the neighborhood of $\Gamma$, is obvious. 
With increasing doping it is the {\it arc intensity} that increases, whereas the {\it arc size} 
barely changes (compare, e.g., Fig.\ref{fig:fs_arc}(c),(d)). Both observations mirror ARPES results and suggest, 
in conjunction with these data here, a simple picture of the underdoped cuprates involving AF and SC clustering concepts. 
To further quantify this behavior, we have calculated the integrated spectral intensity 
in the neighborhood of the FS at different doping levels, $I^{x}_\square$:
\begin{equation}
I^{x}_\square = \int_{-|\omega_D|}^{\mu_F} \int_\square A({\mathbf k},\omega) d\omega d{\mathbf k},
\label{int:spec_weight}
\end{equation}
where $\square$ denotes a window (``window 1(2)'' is a $5^2$ ($7^2$) square) centered around the FS arc midpoint. 
 \begin{figure}
\includegraphics[width=7cm,clip]{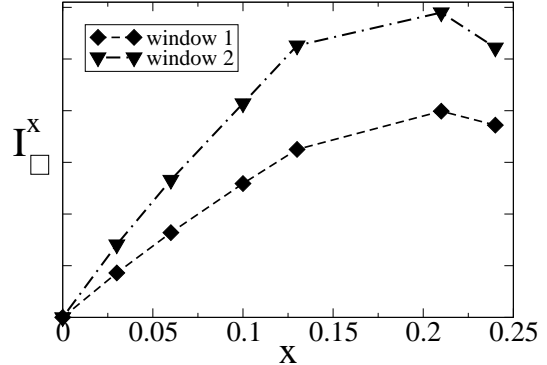}
\caption{\label{fig:int_spec} Integrated spectral weight $I^{x}_\square$, for two different integration windows 
(in $\mathbf k$-space, with frequency cutoff $\omega_{\rm D}$=-0.3) as a function of the hole density $x$. The downturn of $I^{x}_\square$ at large doping is a finite-size effect.}
\end{figure}
Using data gathered from ARPES experiments (however at $\omega_D$$\sim$0), it was shown that this quantity increases approximately 
linearily with doping \cite{re:yoshida03}, which is also what arises from the mean-field calculation (Fig.\ref{fig:int_spec}). This increase is 
almost entirely associated with increased spectral weight around the nodal direction, as is also clear from Fig.\ref{fig:fs_arc}. 
Then, the most natural explanation for the observed behavior of $A$(${\mathbf k}$, $\omega$$\approx$$E_{\rm F}$; $x$) is that it simply reflects the 
increasing amount of $a_{\rm SC}$ rather than the change in particle density (although both are related). 
A linear relationship between $I^x_{\square}$ and the particle density as such 
is not obvious (or even fulfilled) for a clean SC system, but it is naturally explained in a mixed-state picture. 
For the perfect SC ($n_{\rm SC}$$\sim$0.76), the FS is even more clearly defined as in Fig.\ref{fig:fs_arc}(d): 
in this case, the central peaks remain and gain in intensity, whereas 
the small incoherent intensity in the rest of the $k_x$-$k_y$-plane vanishes. Again, this ``sharpening'' of quasiparticle peaks 
as the system becomes more homogeneous - and quasiparticles better defined -  has been observed in ARPES \cite{re:ino00}.  

We also want to mention here that the experimental observation - a doping-independent size of the FS arc - also contradicts 
other popular proposals, which associate the underdoped regime with one of strong pair attraction and significant phase fluctuations: 
In those scenarios, $V$ increases as $x$$\rightarrow$0, but this would entail a FS arc that at the same time is shrinking in {\it size}. 
The simple fact that the size stays constant suggest a relatively doping-independent value of $\Delta$ and, thus, $V$. 
The same is true for scenarios, which rely on additional, non-zero order-parameters (gaps) dominating the underdoped phase: 
the FS should be curtailed and become smaller as this additional order becomes stronger at lower doping, in contrast to what is seen 
experimentally.     

\begin{figure}
\includegraphics[width=8cm,clip]{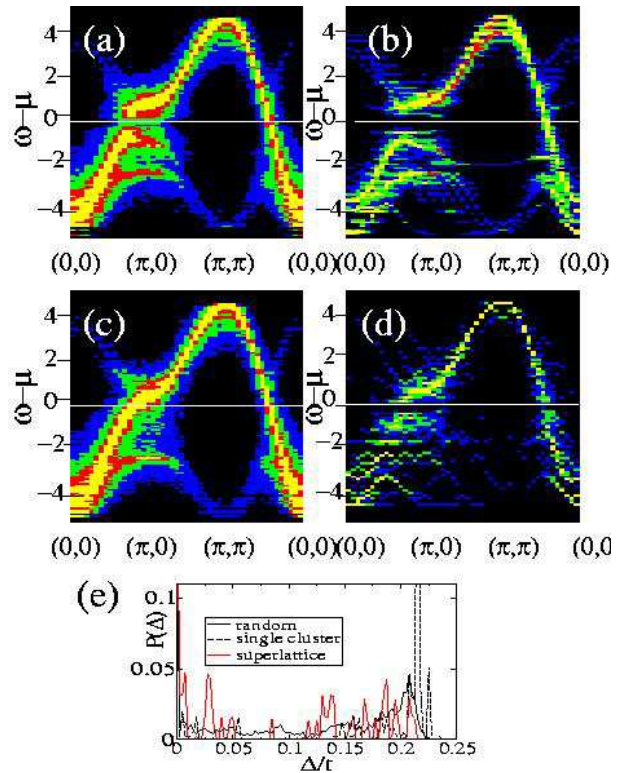}
\caption{\label{fig:arpes_var} $A({\mathbf k}, \omega)$ for doping rate $x$=0.13: (a) same as Fig.\ref{fig:arpes_bdg}(c), 
(b) has one single SC-favoring cluster in the middle, (c) V$_{\mathbf i}$=0 everywhere, even in charge-depleted regions, 
and (d) with SC plaquettes forming a super-lattice. The SC gap distributions $P$($\Delta$) are shown in (e) for 
the random model (a) (thick black line), the single cluster (b) (dashed line) and the superlattice structure in (c) (red line). 
In the first case, the gaps have a Gaussian-like distribution around the maximum, with a broad tail, whereas in case (b) 
there are essentially two peaks only. (c) has a multitude of clearly defined peaks.}
\end{figure}

To lend further insight to our analysis, we have also attempted to a series of different calculations at fixed density 
$\langle$$n$$\rangle$=0.87, but somewhat deviating from the above scenarios: 
(i) first, a model where the random SC configuration as shown in Fig.\ref{fig:definelabel}(a2) is replaced by one with a single cluster, 
occupying the same overall area, in order to better understand the effects of randomness, cluster-size and -shape; (ii) second, a model of 
charge-depleted plaquettes with $V$=0, i.e. a situation of hole-rich and hole-poor phases coexisting, but without the former one 
possessing SC order, and finally (iii) a model, where plaquettes are distributed in an ordered fashion, forming a superlattice 
of 16 plaquettes, each covering a block of 6$\times$6 sites on the usual 32$\times$32 lattice. 
$A$($\mathbf k$, $\omega$)'s resulting from those configurations are 
shown in Fig.\ref{fig:arpes_var}(b)-(d). In the case of (i), the replacement of the random 
configuration by a ``superblock'' leads to a narrowing of the spectral functions, with more clearly defined peaks, 
but otherwise leaves the overall features such as the existence of the two-branches and the FS positions, 
unperturbed. One consequence, however, is the narrowing of the gap distribution function P($\Delta$) 
as the randomness is reduced, reflecting the smaller number of possible environments. 
For the present case, P($\Delta$) consists of a central peak from data well inside the cluster and a smaller peak 
from the SC/AF boundaries, whereas for the totally random case P($\Delta$) can be approximately described by a Gaussian 
distribution, with a multitude of $\Delta$'s. 
This might be of importance, since exactly such a Gaussian distribution has been reported in STM experiments in underdoped BSCCO. 
Then, randomly located SC regions are the best picture to describe the experiments. 
On the other hand, quite drastic changes occur if the plaquettes are chosen as charge-depleted-only regions, 
as shown in Fig.\ref{fig:arpes_var}(c): in such a case a {\it single} band with a clear FS crossing for $\mathbf k$$\sim$($\pi$,0) emerges, 
and such a feature is certainly {\it not} observed in experiments. Therefore, the inclusion 
of a pairing term even in the low-doping limit seems necessary to correctly reproduce the experimental features.

The case of the ordered plaquettes (Fig.\ref{fig:arpes_var}(d)) was motivated by claims that the STM results in underdoped (but SC) BSCCO should be best interpreted 
in terms of charge-ordering \cite{re:vershinin04}; the superlattice structure chosen is the simplest realization of such a CO state, and may help elucidate 
the ARPES measurements of such a complicated, yet highly ordered, phase. Although the overall features are largely unchanged 
when compared to the random cluster situation, the broad peaks observed for the random structure split up into several subpeaks, 
i.e. for some momenta there may be even more than just two solutions. This is even clearer if more traditional representations 
of $A$(${\mathbf k}$, $\omega$) are chosen, not shown here. Although those multiple-peak features may be too weak to be observed 
in current ARPES experiments, and therefore such a state cannot {\it a priori} be ruled out, there is no indication thus far of such a structure being observed, 
at least according to ARPES investigations.        

\section{DOS and $T^*$}
\begin{figure}
\includegraphics[width=8cm,clip]{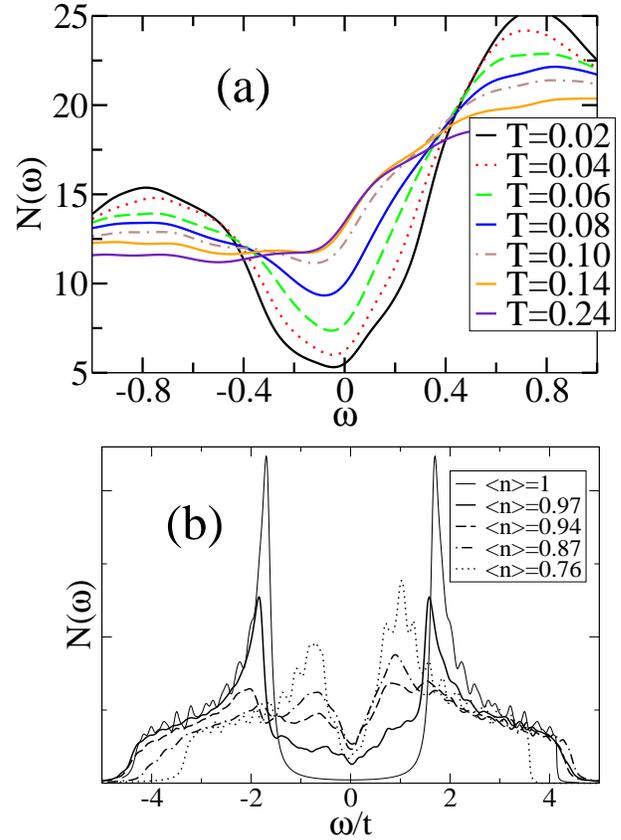}
\caption{\label{fig:dospg} (a) DOS near the Fermi energy (0) for the situation of Fig.~\ref{fig:pdafsc} 
where there are 6 plaquettes on an 8$\times$8 lattice. The disappearence of the PG can be seen as the
temperature is increased. (b) DOS calculated at different densities by mean-field theory at T=0 
for a 32$\times$32 lattice for the disordered model. The clean case ($\langle$$n$$\rangle$=0.76) was calculated for 
a 40$\times$40 lattice to minimize finite-size effects.}
\end{figure}

Calculations of the total DOS for the model with quenched disorder is shown in Fig.~\ref{fig:dospg}a. 
This is at a point in the phase diagram, Fig.~\ref{fig:pdafsc}(b) with 6 impurities, i.e. 
without long-range order either in the AF or the SC sector. The DOS clearly shows a PG that 
disappears at a temperature scale $T^*$. The qualitative physics described here has
already been investigated by the authors in Ref.~\onlinecite{re:alvarez04b} and will not be repeated. 

It is worth noting that in this model the PG is caused by short-range order of either SC or magnetic variables. In other words, if 
the temperature for short
order formation is denoted by $T_{\rm sc}^*$ and $T_{\rm af}^*$ for $d$-wave superconductivity and antiferromagnetism, respectively, 
then the temperature for PG formation is roughly 
$T^*$$\sim$max$(T_{\rm sc}^*,T_{\rm af}^*)$. This explain the dependence of $T^*$ with doping as observed in Fig.~\ref{fig:pdafsc}. 
For low doping the system has strong short- (and
possibly long-range) range AF order and no SC order, therefore, $T^*\sim T^*_{\rm af}$ for low doping. 
As doping increases, the system becomes less and less AF which implies a decrease in  $T^*_{\rm af}$ and 
consequently in  $T^*$. However, as the carrier concentration is increased further, 
the systems starts to present SC order, at short distance first and then at long distances. 
Then, $T^*$ stops decreasing and stays constant or even increases with carrier concentration 
near the optimal doping region.

The BdG equations are less influenced by finite-size effects and, thus, allow for a better understanding of the DOS of a 
mixed AF/SC-state. Although we have only performed $T$$\approx$0-calculations only, some interesting information can be extracted 
from these solutions as discussed below. As the system is doped away from half-filling, the SC regions appear as mid-gap states between the 
original Hubbard bands (Fig.\ref{fig:dospg}(b)), which are gradually filled as doping is increased, at the expense of the original states. 
The associated peaks in $N$($\omega$) gain in strength, whereas the energy states belonging 
to the AF sites gradually fade away, giving way to the usual DOS of a dSC. From the most naive point of view, this two-peak 
structure in the DOS can be associated with the PG, wherein it is the disappearance of the AF peak (at higher energies) 
- which should be happening at ever lower T's as $x$ is increased - that determines $T^\star$($x$). But there are also some other subtle issues that need 
to be considered following Fig.\ref{fig:dospg}(b): the SC peak seems to be travelling towards higher energy, rather than towards 
$E_{\rm F}$. This can already be seen in the corresponding ARPES data, but is much more obvious in the DOS. This is in disagreement with 
data from ARPES, which clearly show the SC band moving closer to $E_{\rm F}$ as $x$ is increased. In this framework, the PG is dictated 
by the energy position of the charge-depleted phase in the presence of the half-filled insulator. This is an important, yet subtle 
issue since the energies involved are rather small; yet it should be explored in further considerations of the mixed-state picture. 
Summarizing, although the description of PG physics is not perfect, at least the key features (such as its existence) 
are neatly reproduced by our mixed-phase model.

\section{Local DOS}

\begin{figure}
\includegraphics[width=8cm,clip]{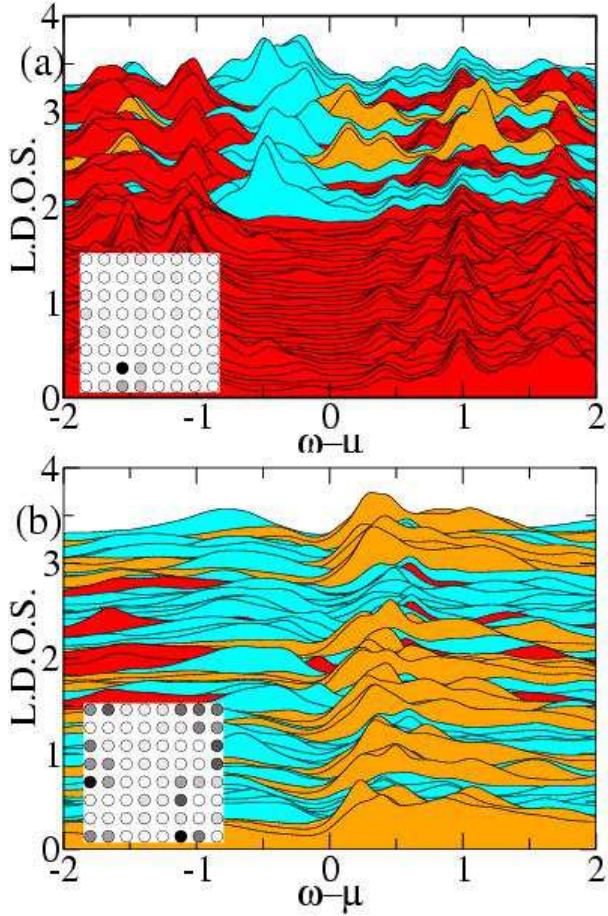}
\caption{\label{fig:ldos8x8a}  Local DOS $N_{\bf i}(\omega)$ vs. $\omega-\mu$ on an 8$\times$8 lattice
with $1$ and $6$ ``plaquettes'', respectively, plotted in ``linear order''. The color 
convention is as follows: red indicates AF sites; orange, SC sites and 
cyan, metallic sites. The inset shows the $d$-wave gap distribution, 
$|\Delta_{\bf i,x}| (\cos(\phi_{\bf i})-\cos(\phi_{\bf i+x}))$
 in each case.}
\end{figure}
\begin{figure}
\includegraphics[width=8cm,clip]{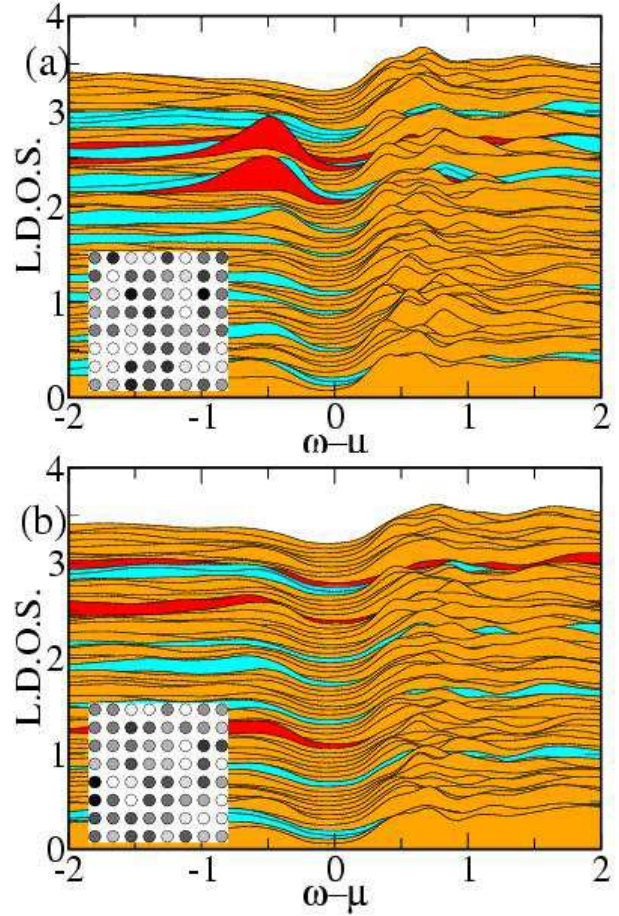}
\caption{\label{fig:ldos8x8b}  $N_{\bf i}(\omega)$ on an 8$\times$8 lattice
with $20$ and $24$ ``plaquettes'', respectively plotted in ``linear order''. 
Color convention as in Fig.\ref{fig:ldos8x8a}. Just as in Fig.\ref{fig:ldos8x8a}, the 
datasets for each site ${\bf i}$ are offset by a small amount from each other 
to allow for better visibility.}
\end{figure} 

\begin{figure}
\includegraphics[width=8.7cm,clip]{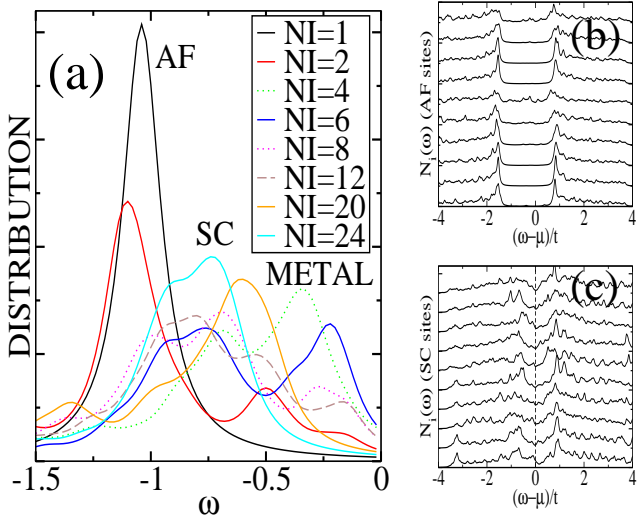}
\caption{\label{fig:mld8x8}Distribution of the local DOS vs. $\omega-\mu$ on 
an 8$\times$8 lattice with the concentration of ``plaquettes'' shown and parameters 
corresponding to Figs.~\ref{fig:ldos8x8a}-\ref{fig:ldos8x8b}. NI stands for the number of impurities.
(b) shows $N_{\bf i}$($\omega$) for sites, which are predominantly AF, whereas (c) those for SC ordered sites. 
Those data were taken on a disordered lattice with 28$\times$28 sites, and about 60$\%$ SC. The parameters 
were $U_{\bf i}$=4 for the AF regions, and $V_{\bf i}$=-1 for the SC sites.}
\end{figure}

Consider again a system with inhomogeneous couplings $J_{\bf i}$ and $V_{\bf i}$ in a such a 
way as to produce AF and SC regions in the sample. We calculate $N_{\bf i}(\omega)$ 
for all sites, displaying them in ``linear order''
\footnote{``Linear order'' is the order defined in the following way: $(x,y)<(x',y')$ if and 
only if $x+yL<x'+y'L$ with $L$ the lattice length.} in Fig.~\ref{fig:ldos8x8a}-\ref{fig:ldos8x8b}, 
for various plaquettes concentrations as indicated. The results show that there are two types 
of curves or ``modes'', and that each site contributes to a different ``mode'' of the LDOS. 
For example, all AF sites present a clear gap
(Fig.~\ref{fig:ldos8x8a}a), whereas SC sites have the equivalent of a $d-$wave gap 
(on a finite-size system). This seems in agreement with STM
measurements\cite{re:lang02} in  underdoped Bi$_2$Sr$_2$CaCu$_2$O$_{8+\delta}$.
For example, Figure 3 of Ref.~\onlinecite{re:lang02} shows the differential conductance along a path on the
sample vs. the bias, indicating two types of regions - similarly to what is observed 
in Fig.~\ref{fig:ldos8x8a}b and Fig.~\ref{fig:ldos8x8b}a of the present work, which 
represent intermediate states which do not present a dominant global order parameter 
(as is the case for Figs.\ref{fig:ldos8x8a}(a), \ref{fig:ldos8x8b}(a). 
Certainly these data still suffer from finite-size problems, but the basic issues are captured nevertheless. 
Also, note that the transition from the SC to the AF regions is gradual with an intermediate phase 
that has a small SC-like gap, but lacks the pronounced (coherence) peak of the SC state.
These results favor the idea that ``exotic'' phases are not needed to explain the nature 
of underdoped cuprates, but instead that quenched disorder creates 
regions of local SC or AF order when those two phases strongly compete. 

The distribution of intensities of the LDOS for each frequency range is shown in Fig.~\ref{fig:mld8x8}. 
The AF, SC and metallic (non-SC) contributions are present with different
intensities depending on the concentration of SC plaquettes as shown. 
A metallic, but not SC, phase appears due to the value of the chemical
potential used around each plaquette. This detail is not crucial to obtain the data presented here, 
but instead provides a more realistic separation between SC and
AF regions. With a simpler model for the plaquettes, the metallic peak would not be present.

This is even more obvious in the mean-field approximation, where sites with local AF or SC order can be easily 
identified. Travelling along certain cuts in a mixed-phase system, 
which show either AF or SC order, and measuring $N_{\bf i}$($\omega$) along this path (see Fig.\ref{fig:mld8x8}(b),(c)), 
it is revealed that the signals are vastly different in either case. Whereas the AF sites (derived from the value of $m_{\bf i}$ 
have a clear and well-defined gap, the SC sites resemble those of a $d$-wave SC. From these results it is also clear that there 
is not just one SC gap (which may, for example, be extracted from the position of the first main peak), 
but rather a distribution of gap values, presumably reflecting the random nature of the samples, 
as depicted in Fig.\ref{fig:definelabel}(b)-(b3). 
As the hole doping level is increased, the relative amount of SC data {\it taken ``in the bulk''} increases, 
leading to a narrowing of the gap distribution.    
Those data, for example, should be compared with those for Ca$_{1.92}$Na$_{0.08}$CuO$_2$Cl$_2$ \cite{Kohsaka04}, 
where similar features were observed; sites with a large excitation gap were found to be 
insulating, whereas the other ones were metallic. As the number of charge-carriers is modified, 
the relative amount of those two phases changes, whereas the dI/dV-signals themselves remain fairly 
independent of the doping level.

\section{Conclusions}
In a recent publication, 
a phenomenological model for cuprates was introduced \cite{re:alvarez04b}. This model 
includes itinerant fermions coupled to classical degrees of freedom that represent the
AF and SC order parameters. The model can be studied integrating out the fermions
and Monte Carlo simulating the classical fields.
The simple characteristics of this model allow to investigate the crossover
from AF to SC, with or without quenched disorder incorporated, and regardless of whether
homogeneous or inhomogeneous states emerge as ground states. The simplified character
of the model, as compared with the much more difficult to study 
Hubbard model, allows for numerical studies at any
electronic density and temperature, and the evaluation of dynamical properties as well.
Due to present day limitations in the analysis of many-body problems,
{\it the complex physics of transition metal oxides at nanometer length scales can only
be captured with phenomenological models}, such as those recently discussed 
by Alvarez {\it et al.}\cite{re:alvarez04b}.

In the present publication, the previous effort has been extended to the analysis
of photoemission one-particle spectral functions and the local DOS, 
comparing theory with experiments. It has been observed that without quenched disorder (clean limit)
the model presents various phases: SC, AF, 
local coexistence of SC and AF orders,
and striped states. However,
the spectral density calculated for all these states does not reproduce the
experimental measurements for cuprate superconductors as reported 
in Ref.~\onlinecite{re:yoshida03}. 
To make progress in the theoretical description of experiments, quenched disorder was added
to the system, inducing regions of SC and
AF order, interpolating between the two fairly uniform AF and SC states.
In this case, the ARPES spectral weight presents $two$ branches, one induced
by the AF background and the other by the
SC regions or islands. This spectral weight shows clear similarities with
the experimental observations and, therefore, the ``nodal'' regions observed
near the Fermi energy for underdoped compounds are explained within the context 
of our model as induced by the SC regions.

The calculation of the DOS revealed a PG for the regions of the phase diagram 
where no long range order dominates, although local order exists. 
The disappearence of the PG with
temperature gave an estimation of the temperature scale, $T^*$. Moreover, the LDOS 
calculated in the inhomogeneous system presents two types of curves or ``modes'', 
corresponding to AF and SC clusters. 
This observation is in agreement with STM measurements\cite{re:lang02}.

In summary, a simple model is able to capture several features found experimentally in
high temperature superconductors. In this theory, the glassy state is formed by
patches of both phases, as already discussed in Ref.\onlinecite{re:alvarez04b}, and
it can present giant responses to small external perturbations. In this work, it was
shown that ARPES, DOS and LDOS experimental information in the underdoped regime
can also be rationalized using
the same simple theory. Our model does not address directly the important issue of
the origin of pairing in the SC state, but can describe phenomenologically the
AF vs. SC competition. The results support the view that underdoped cuprate superconductors 
are inhomogeneous at the nanoscale, and that only well-known competing states (AF and SC, 
perhaps complemented by stripes) are needed to understand this very mysterious regime of the cuprates.

\begin{acknowledgments}
This research was performed in part at the Oak Ridge National Laboratory, managed by UT-Battelle, LLC, 
for the U.S. Department of Energy under Contract DE-AC05-00OR22725.
E.D. is supported by NSF through the grant DMR-0443144. We are indebted 
to David S\'en\'echal for suggestions about the presentation of spectral density data.
\end{acknowledgments}

\appendix*
\section{Diagonalization Procedure}

To diagonalize Eq.(\ref{eq:hamfermi0}) a modified Bogoliubov transformation\cite{re:ghosal02} 
needs to be applied. After some algebra it can be shown that Eq.~(\ref{eq:hamfermi0}) becomes:
\begin{eqnarray}
H &= & \sum_{n,n'=1}^{N}\sum_{\alpha,\alpha'=0}^{1}\sum_{\langle \mathbf i\mathbf j \rangle}
 a^*_{n+\alpha N}(\mathbf i) H_{\mathbf i\alpha ,\mathbf j\alpha' }^\uparrow a_{n'+\alpha' N}(\mathbf j)
\gamma_{n\uparrow}^\dagger\gamma_{n'\uparrow}\nonumber\\
&-&
\sum_{n,n'=1}^{N}\sum_{\alpha,\alpha'=0}^{1}\sum_{\langle \mathbf i\mathbf j \rangle} b_{n+\alpha N}^*(\mathbf i) 
H_{\mathbf i\alpha,\mathbf j\alpha'}^\downarrow
b_{n'+\alpha' N}(\mathbf j)\gamma_{n'\downarrow}\gamma_{n\downarrow}^\dagger+\nonumber\\
&+&
\sum_{n=1}^{N}M_n +
\frac {1}{2}\sum_{{\bf i},\alpha}\frac {1}{V_{\bf i}}|\Delta_{{\bf i}\alpha}|^2,
\end{eqnarray}
where $H^\uparrow$ and $H^\downarrow$ are the $2N\times2N$ matrices given by
\begin{equation}
H^\uparrow=\left(
\begin{tabular}{ll}
$\hat{K}+\hat{J}$ & \hspace*{6mm}$\hat{\Delta}$ \\
\hspace*{4mm}$\hat{\Delta}^*$ & $-\hat{K}+\hat{J}$\\
\end{tabular}
\right),
\label{eq:hamup}
\end{equation}

\begin{equation}
H^\downarrow=\left(
\begin{tabular}{ll}
$\hat{K}-\hat{J}$ & \hspace*{6mm}$\hat{\Delta}$ \\
\hspace*{4mm}$\hat{\Delta}^*$ & $-\hat{K}-\hat{J}$\\
\end{tabular}
\right).
\label{eq:hamdown}
\end{equation} 
$\hat{K}$ ($\hat{\Delta}$) is a $N\times N$ matrix, such that $\hat{K}_{{\rm \mathbf i}{\rm \mathbf j}}= -t$ 
($\hat{\Delta}_{{\rm \mathbf i}{\rm \mathbf j}}=\Delta_{{\mathbf i},\alpha}$) only if ${\rm \mathbf i}$ and ${\rm \mathbf j}$
are n.n. sites and 0 otherwise. 
$\hat{J}_{\mathbf i\mathbf j}=J_{\mathbf i} S_{\mathbf i}^z \delta_{\mathbf i\mathbf j}$ and
$M_n= \sum_{\mathbf i\mathbf j}b_n^*(\mathbf i) (\hat{K}_{\mathbf i\mathbf j}-\hat{J}_{\mathbf i\mathbf j}) b_n(\mathbf j)+ a_{n+N}^*(\mathbf i) (\hat{K}_{\mathbf i\mathbf j}-\hat{J}_{\mathbf i\mathbf j}) a_{n+N}(\mathbf j)$.
The c-numbers $a_n(\mathbf i)$ ($b_n(\mathbf i)$) are chosen, so that they diagonalize $H^\uparrow$ ($H^\downarrow$), i.e.:
\begin{equation}
\sum_{\mathbf i\mathbf j}a_{n+\alpha N} (\mathbf i) H^\uparrow_{\mathbf i\alpha,\mathbf j\alpha'} 
a_{n'+\alpha' N} (\mathbf j) = E^\uparrow_{n+\alpha N} \delta_{n+\alpha N,n'+\alpha' N}. 
\label{eq:uparrowtransform}
\end{equation}
Then, the total energy can be written as 
\begin{eqnarray}
E_{\rm total}&=&\sum_{n=1}^{n=N} (E_n^\uparrow f(\beta E_n^\uparrow) + E_n^\downarrow f(\beta E_n^\downarrow))\nonumber\\
&+&\sum_{n=1}^{N}(M_n-E_n^\downarrow)+\frac {1}{2}\sum_{{\bf i},\alpha}\frac {1}{V_{\bf i}}|\Delta_{{\bf i}\alpha}|^2,
\label{eq:etotal}
\end{eqnarray}
where $f(x)=1/(1+{\rm e}^x)$ is the Fermi function.
The sum is only over the $N$ largest eigenvalues of $H^\uparrow$, $\{E_n^\uparrow\}_{1\le n\le N}$, and the $N$ largest eigenvalues of $H^\downarrow$,
$\{E_n^\downarrow\}_{1\le n\le N}$. This expression
involves the eigenvalues of both matrices. Alternatively, it is possible to express $E_{\rm total}$ in terms of the 
eigenvalues of only one matrix, $H^\uparrow$ for example, which is the more efficient way for the MC simulation.
 To understand that, first note that the eigenvalues of $H^\downarrow$ have the opposite signs as  those
of $H^\uparrow$. Let:
\begin{equation} 
S=\left(
\begin{tabular}{ll}
$0$ & $\hat{I}$ \\
$\hat{I}$ & $0$\\
\end{tabular}
\right),
\end{equation}
where $\hat{I}$ is the $N\times N$ identity matrix.
Then $S=S^{-1}$ and $H^\uparrow= -S H^\downarrow S^{-1}$. Therefore, if $|\nu\rangle$ is an eigenvector of $H^\uparrow$ with eigenvalue $E^\uparrow_n$, then
$S^{-1}|\nu\rangle$ is an eigenvector of $H^\downarrow$ with eigenvalue $-E_n^\uparrow$. 
This proves that the eigenvalues of $H^\downarrow$ and
$H^\uparrow$ are the opposites of one another. From this discussion it also follows immediately
  that if $a$ is an eigenvector of $H^\uparrow$, then
$b$ defined by $b_n=a_{n+N},\,b_{n+N}=a_n$ $\forall n\in{[}1,N{]}$ is an eigenvector of $H^\downarrow$ so the eigenvectors of
one matrix can be obtained from the eigenvectors of the other.
Now, let $\{E_n^\uparrow\}_{N+1\le n\le 2N}$ be the $N$ lowest eigenvalues of $H^\uparrow$ and
similarly   $\{E_n^\downarrow\}_{N+1\le n\le 2N}$ the $N$ lowest eigenvalues of $H^\downarrow$. Therefore, $E^\downarrow_n= -E^\uparrow_{n+N}$ $\forall n\in{[}1,N{]}$.
Then, the second term in Eq.~(\ref{eq:etotal}) becomes $-E^\uparrow_{n+N} f(-\beta E_{n+N}^\uparrow)$ and using that $f(-x)=1-f(x)$:
\begin{equation}
E_{\rm total}=\sum_{n=1}^{n=2N}E_n^\uparrow f(\beta E_n^\uparrow)+\sum_{n=1}^{N}M_n+
\frac {1}{2}\sum_{{\bf i},\alpha}\frac {1}{V_{\bf i}}|\Delta_{{\bf i}\alpha}|^2
,
\end{equation}
and this  expression was used in the MC evolution of the system.

Most observables are calculated by replacing the electron operator $c_{\mathbf i\sigma}$ by Bogoliubov operators via
Eq.~(\ref{eq:bogoliubov}). 
For example, the number of particles, $N_e$, is given by the average of
 $\sum_{\mathbf i}c_{\mathbf i\sigma}^\dagger c_{\mathbf i\sigma}$ and in
terms of $a_n(\mathbf i)$ its expression is:
\begin{equation}
N_e=\sum_{n=1}^{n=2N} |a_n|^2 + 2\sum_{n=1}^N(|a_n|^2-|a_{n+N}|^2)(f(\beta E^\uparrow_n)+f(\beta E^\downarrow_{n})),
\label{eq:ne}
\end{equation}
where we have used the abreviation $|a_n|^2=\sum_{\mathbf i} |a_n(\mathbf i)|^2$.
Note that unlike the standard Bogoliubov expression for the number of electrons, the second term in Eq.~(\ref{eq:ne}) can contribute
even at $T=0$ due to the fact that $E^\uparrow_n$ can be negative for $J$ finite. Moreover, unlike standard spin-fermion models, such as
those for manganites and cuprates,\cite{re:moraghebi01} the
number of electrons depends not only on the eigenvalues of the one-particle sector, but also on the eigenvectors.

As a particular case consider $A({\mathbf r},t)$, which is defined by the expression:
\begin{equation}
A(\mathbf r,t)=<\sum_{\mathbf l} c^\dagger_{\mathbf l\sigma}(t) c_{\mathbf l+\mathbf r,\sigma}(0) + H.c.>,
\label{eq:art}
\end{equation}
where $\langle$$\hdots$$\rangle$ denotes thermal averaging.
Applying the modified BdG transformation, Eq.~(\ref{eq:bogoliubov}), Eq.~(\ref{eq:art}) is calculated using:
\begin{equation}
A(\mathbf r,\omega)=\sum_n X_{n}(\mathbf r) \delta(\omega-E_n^\uparrow)+
Y_{n}(\mathbf r) \delta(\omega+E_n^\uparrow),
\label{eq:akw2}
\end{equation}
where
\begin{equation}
X_{n}(\mathbf r)=\sum_{\mathbf l} a_{n}^*(\mathbf l) a_{n}(\mathbf l+\mathbf r),
\end{equation}
and a similar expression is valid for $Y_{n}$.
Eq.~(\ref{eq:akw2}) can be Fourier-transformed to obtain $A(\mathbf k,\omega)$, but it is faster to do that after performing the average, and
that route has been followed in the present work.

In STM experiments, one typically measures the change of a (local) tunneling current d$I$/d$V$, a quantity, which - under suitable assumptions - 
is proportional to $N(\mathbf i,\omega)$ and thus allows for a direct mapping of the local electronic states. $N(\mathbf i,\omega)$ is but 
the Fourier-transform of  $A(\mathbf k,\omega)$ and can be just as easily evaluated:
\begin{equation}
N(\mathbf i,\omega)=\sum_{n=1}^{N} |a_{n}(\mathbf i)|^2 \delta(\omega-E_n)+
|a_{n+N}(\mathbf i)|^2 \delta(\omega+E_n).
\end{equation}

\vspace*{1cm}$^\star$ present address: Department of Physics and Astronomy, The University of Tennessee, Knoxville,
Tennessee 37996
\bibliography{thesis}

\begin{thebibliography}{39}
\expandafter\ifx\csname natexlab\endcsname\relax\def\natexlab#1{#1}\fi
\expandafter\ifx\csname bibnamefont\endcsname\relax
  \def\bibnamefont#1{#1}\fi
\expandafter\ifx\csname bibfnamefont\endcsname\relax
  \def\bibfnamefont#1{#1}\fi
\expandafter\ifx\csname citenamefont\endcsname\relax
  \def\citenamefont#1{#1}\fi
\expandafter\ifx\csname url\endcsname\relax
  \def\url#1{\texttt{#1}}\fi
\expandafter\ifx\csname urlprefix\endcsname\relax\def\urlprefix{URL }\fi
\providecommand{\bibinfo}[2]{#2}
\providecommand{\eprint}[2][]{\url{#2}}

\bibitem[{\citenamefont{Vojta}(2002)}]{re:vojta02}
\bibinfo{author}{\bibfnamefont{M.}~\bibnamefont{Vojta}},
  \bibinfo{journal}{Phys. Rev. B} \textbf{\bibinfo{volume}{66}},
  \bibinfo{pages}{104505} (\bibinfo{year}{2002}).

\bibitem[{\citenamefont{Nayak}(2000)}]{re:nayak00}
\bibinfo{author}{\bibfnamefont{C.}~\bibnamefont{Nayak}},
  \bibinfo{journal}{Phys. Rev. B} \textbf{\bibinfo{volume}{62}},
  \bibinfo{pages}{4880} (\bibinfo{year}{2000}).

\bibitem[{\citenamefont{Wen and Lee}(1996)}]{re:wen96}
\bibinfo{author}{\bibfnamefont{X.-G.} \bibnamefont{Wen}} \bibnamefont{and}
  \bibinfo{author}{\bibfnamefont{P.}~\bibnamefont{Lee}},
  \bibinfo{journal}{Phys. Rev. Lett.} \textbf{\bibinfo{volume}{76}},
  \bibinfo{pages}{503} (\bibinfo{year}{1996}).

\bibitem[{\citenamefont{Affleck and Marston}(1987)}]{re:affleck87}
\bibinfo{author}{\bibfnamefont{I.}~\bibnamefont{Affleck}} \bibnamefont{and}
  \bibinfo{author}{\bibfnamefont{J.}~\bibnamefont{Marston}},
  \bibinfo{journal}{Phys. Rev. B} \textbf{\bibinfo{volume}{37}},
  \bibinfo{pages}{8865} (\bibinfo{year}{1987}).

\bibitem[{\citenamefont{Varma}(1999)}]{re:varma99}
\bibinfo{author}{\bibfnamefont{C.}~\bibnamefont{Varma}},
  \bibinfo{journal}{Phys. Rev. Lett.} \textbf{\bibinfo{volume}{83}},
  \bibinfo{pages}{3538} (\bibinfo{year}{1999}).

\bibitem[{\citenamefont{Chakravarty et~al.}(2001)\citenamefont{Chakravarty,
  Laughlin, Morr, and Nayak}}]{re:chakravarty01}
\bibinfo{author}{\bibfnamefont{S.}~\bibnamefont{Chakravarty}},
  \bibinfo{author}{\bibfnamefont{R.}~\bibnamefont{Laughlin}},
  \bibinfo{author}{\bibfnamefont{D.}~\bibnamefont{Morr}}, \bibnamefont{and}
  \bibinfo{author}{\bibfnamefont{C.}~\bibnamefont{Nayak}},
  \bibinfo{journal}{Phys. Rev. B} \textbf{\bibinfo{volume}{63}},
  \bibinfo{pages}{094503} (\bibinfo{year}{2001}).

\bibitem[{\citenamefont{Emery and Kivelson}(1995)}]{re:emery95}
\bibinfo{author}{\bibfnamefont{V.}~\bibnamefont{Emery}} \bibnamefont{and}
  \bibinfo{author}{\bibfnamefont{S.}~\bibnamefont{Kivelson}},
  \bibinfo{journal}{Nature} \textbf{\bibinfo{volume}{374}},
  \bibinfo{pages}{434} (\bibinfo{year}{1995}).

\bibitem[{\citenamefont{Hunt et~al.}(2001)\citenamefont{Hunt, Singer,
  Cederstrom, and Imai}}]{re:hunt01}
\bibinfo{author}{\bibfnamefont{A.~W.} \bibnamefont{Hunt}},
  \bibinfo{author}{\bibfnamefont{P.~M.} \bibnamefont{Singer}},
  \bibinfo{author}{\bibfnamefont{A.~F.} \bibnamefont{Cederstrom}},
  \bibnamefont{and} \bibinfo{author}{\bibfnamefont{T.}~\bibnamefont{Imai}},
  \bibinfo{journal}{Phys. Rev. B} \textbf{\bibinfo{volume}{64}},
  \bibinfo{pages}{134525} (\bibinfo{year}{2001}).

\bibitem[{\citenamefont{Damascelli et~al.}(2003)\citenamefont{Damascelli,
  Hussain, and Shen}}]{re:damascelli03}
\bibinfo{author}{\bibfnamefont{A.}~\bibnamefont{Damascelli}},
  \bibinfo{author}{\bibfnamefont{Z.}~\bibnamefont{Hussain}}, \bibnamefont{and}
  \bibinfo{author}{\bibfnamefont{Z.-X.} \bibnamefont{Shen}},
  \bibinfo{journal}{Rev. Mod. Phys.} \textbf{\bibinfo{volume}{75}},
  \bibinfo{pages}{473} (\bibinfo{year}{2003}).

\bibitem[{\citenamefont{Ino et~al.}(2000)\citenamefont{Ino, Kim, Nakamura,
  Yoshida, Mizokawa, Shen, Fujimori, Kakeshita, Eisaki, and Uchida}}]{re:ino00}
\bibinfo{author}{\bibfnamefont{A.}~\bibnamefont{Ino}},
  \bibinfo{author}{\bibfnamefont{C.}~\bibnamefont{Kim}},
  \bibinfo{author}{\bibfnamefont{M.}~\bibnamefont{Nakamura}},
  \bibinfo{author}{\bibfnamefont{T.}~\bibnamefont{Yoshida}},
  \bibinfo{author}{\bibfnamefont{T.}~\bibnamefont{Mizokawa}},
  \bibinfo{author}{\bibfnamefont{Z.-X.} \bibnamefont{Shen}},
  \bibinfo{author}{\bibfnamefont{A.}~\bibnamefont{Fujimori}},
  \bibinfo{author}{\bibfnamefont{T.}~\bibnamefont{Kakeshita}},
  \bibinfo{author}{\bibfnamefont{H.}~\bibnamefont{Eisaki}}, \bibnamefont{and}
  \bibinfo{author}{\bibfnamefont{S.}~\bibnamefont{Uchida}},
  \bibinfo{journal}{Phys. Rev. B} \textbf{\bibinfo{volume}{62}},
  \bibinfo{pages}{4137} (\bibinfo{year}{2000}).

\bibitem[{\citenamefont{Yoshida et~al.}(2003)\citenamefont{Yoshida, Zhou,
  Sasagawa, Yang, Bogdanov, Lanzara, Hussain, Mizokawa, Fujimori, Eisaki
  et~al.}}]{re:yoshida03}
\bibinfo{author}{\bibfnamefont{T.}~\bibnamefont{Yoshida}},
  \bibinfo{author}{\bibfnamefont{X.~J.} \bibnamefont{Zhou}},
  \bibinfo{author}{\bibfnamefont{T.}~\bibnamefont{Sasagawa}},
  \bibinfo{author}{\bibfnamefont{W.~L.} \bibnamefont{Yang}},
  \bibinfo{author}{\bibfnamefont{P.~V.} \bibnamefont{Bogdanov}},
  \bibinfo{author}{\bibfnamefont{A.}~\bibnamefont{Lanzara}},
  \bibinfo{author}{\bibfnamefont{Z.}~\bibnamefont{Hussain}},
  \bibinfo{author}{\bibfnamefont{T.}~\bibnamefont{Mizokawa}},
  \bibinfo{author}{\bibfnamefont{A.}~\bibnamefont{Fujimori}},
  \bibinfo{author}{\bibfnamefont{H.}~\bibnamefont{Eisaki}},
  \bibnamefont{et~al.}, \bibinfo{journal}{Phys. Rev. Lett.}
  \textbf{\bibinfo{volume}{91}}, \bibinfo{pages}{027001}
  (\bibinfo{year}{2003}).

\bibitem[{\citenamefont{Alvarez et~al.}(2005)\citenamefont{Alvarez, Mayr,
  Moreo, and Dagotto}}]{re:alvarez04b}
\bibinfo{author}{\bibfnamefont{G.}~\bibnamefont{Alvarez}},
  \bibinfo{author}{\bibfnamefont{M.}~\bibnamefont{Mayr}},
  \bibinfo{author}{\bibfnamefont{A.}~\bibnamefont{Moreo}}, \bibnamefont{and}
  \bibinfo{author}{\bibfnamefont{E.}~\bibnamefont{Dagotto}},
  \bibinfo{journal}{Phys. Rev. B} \textbf{\bibinfo{volume}{71}},
  \bibinfo{pages}{014514} (\bibinfo{year}{2005}).

\bibitem[{\citenamefont{Bozovic et~al.}(2004)\citenamefont{Bozovic, Logvenov,
  Verhoeven, Caputo, Goldobin, and Beasley}}]{re:bozovic04}
\bibinfo{author}{\bibfnamefont{I.}~\bibnamefont{Bozovic}},
  \bibinfo{author}{\bibfnamefont{G.}~\bibnamefont{Logvenov}},
  \bibinfo{author}{\bibfnamefont{M.~A.~J.} \bibnamefont{Verhoeven}},
  \bibinfo{author}{\bibfnamefont{P.}~\bibnamefont{Caputo}},
  \bibinfo{author}{\bibfnamefont{E.}~\bibnamefont{Goldobin}}, \bibnamefont{and}
  \bibinfo{author}{\bibfnamefont{M.~R.} \bibnamefont{Beasley}},
  \bibinfo{journal}{Phys. Rev. Lett.} \textbf{\bibinfo{volume}{93}},
  \bibinfo{pages}{157002} (\bibinfo{year}{2004}).

\bibitem[{\citenamefont{Decca et~al.}(2000)\citenamefont{Decca, Drew,
  Osquiguil, Maiorov, and Guimpel}}]{re:decca00}
\bibinfo{author}{\bibfnamefont{R.~S.} \bibnamefont{Decca}},
  \bibinfo{author}{\bibfnamefont{H.~D.} \bibnamefont{Drew}},
  \bibinfo{author}{\bibfnamefont{E.}~\bibnamefont{Osquiguil}},
  \bibinfo{author}{\bibfnamefont{B.}~\bibnamefont{Maiorov}}, \bibnamefont{and}
  \bibinfo{author}{\bibfnamefont{J.}~\bibnamefont{Guimpel}},
  \bibinfo{journal}{Phys. Rev. Lett.} \textbf{\bibinfo{volume}{85}},
  \bibinfo{pages}{3708} (\bibinfo{year}{2000}).

\bibitem[{\citenamefont{J.~Quintanilla and Capelle}(2003)}]{re:quintanilla03}
\bibinfo{author}{\bibfnamefont{L.~O.} \bibnamefont{J.~Quintanilla}}
  \bibnamefont{and} \bibinfo{author}{\bibfnamefont{K.}~\bibnamefont{Capelle}},
  \bibinfo{journal}{Phys. Rev. Lett.} \textbf{\bibinfo{volume}{90}},
  \bibinfo{pages}{089703} (\bibinfo{year}{2003}).

\bibitem[{\citenamefont{Dagotto et~al.}(2001)\citenamefont{Dagotto, Hotta, and
  Moreo}}]{re:dagotto01}
\bibinfo{author}{\bibfnamefont{E.}~\bibnamefont{Dagotto}},
  \bibinfo{author}{\bibfnamefont{T.}~\bibnamefont{Hotta}}, \bibnamefont{and}
  \bibinfo{author}{\bibfnamefont{A.}~\bibnamefont{Moreo}},
  \bibinfo{journal}{Physics Reports} \textbf{\bibinfo{volume}{344}},
  \bibinfo{pages}{1} (\bibinfo{year}{2001}).

\bibitem[{\citenamefont{Dagotto}(2002)}]{re:dagotto02}
\bibinfo{editor}{\bibfnamefont{E.}~\bibnamefont{Dagotto}}, ed.,
  \emph{\bibinfo{title}{Nanoscale Phase Separation and Colossal
  Magnetoresistance}} (\bibinfo{publisher}{Springer Verlag},
  \bibinfo{address}{Berlin}, \bibinfo{year}{2002}).

\bibitem[{\citenamefont{Burgy et~al.}(2001)\citenamefont{Burgy, Mayr,
  Martin-Mayor, Moreo, and Dagotto}}]{re:burgy01}
\bibinfo{author}{\bibfnamefont{J.}~\bibnamefont{Burgy}},
  \bibinfo{author}{\bibfnamefont{M.}~\bibnamefont{Mayr}},
  \bibinfo{author}{\bibfnamefont{V.}~\bibnamefont{Martin-Mayor}},
  \bibinfo{author}{\bibfnamefont{A.}~\bibnamefont{Moreo}}, \bibnamefont{and}
  \bibinfo{author}{\bibfnamefont{E.}~\bibnamefont{Dagotto}},
  \bibinfo{journal}{Phys. Rev. Lett.} \textbf{\bibinfo{volume}{87}},
  \bibinfo{pages}{277202} (\bibinfo{year}{2001}).

\bibitem[{\citenamefont{Zhang}(1997)}]{re:zhang97}
\bibinfo{author}{\bibfnamefont{S.}~\bibnamefont{Zhang}},
  \bibinfo{journal}{Science} \textbf{\bibinfo{volume}{275}},
  \bibinfo{pages}{1089} (\bibinfo{year}{1997}).

\bibitem[{\citenamefont{Chen et~al.}(2004)\citenamefont{Chen, Capponi, Alet,
  and Zhang}}]{re:chen03}
\bibinfo{author}{\bibfnamefont{H.-D.} \bibnamefont{Chen}},
  \bibinfo{author}{\bibfnamefont{S.}~\bibnamefont{Capponi}},
  \bibinfo{author}{\bibfnamefont{F.}~\bibnamefont{Alet}}, \bibnamefont{and}
  \bibinfo{author}{\bibfnamefont{S.-C.} \bibnamefont{Zhang}},
  \bibinfo{journal}{Phys. Rev. B} \textbf{\bibinfo{volume}{70}},
  \bibinfo{pages}{024516} (\bibinfo{year}{2004}).

\bibitem[{\citenamefont{Lang et~al.}(2002)\citenamefont{Lang, Madhavan,
  Hoffman, Hudson, Eisaki, Uchida, and Davis}}]{re:lang02}
\bibinfo{author}{\bibfnamefont{K.}~\bibnamefont{Lang}},
  \bibinfo{author}{\bibfnamefont{V.}~\bibnamefont{Madhavan}},
  \bibinfo{author}{\bibfnamefont{J.~E.} \bibnamefont{Hoffman}},
  \bibinfo{author}{\bibfnamefont{E.~W.} \bibnamefont{Hudson}},
  \bibinfo{author}{\bibfnamefont{H.}~\bibnamefont{Eisaki}},
  \bibinfo{author}{\bibfnamefont{S.}~\bibnamefont{Uchida}}, \bibnamefont{and}
  \bibinfo{author}{\bibfnamefont{J.~C.} \bibnamefont{Davis}},
  \bibinfo{journal}{Nature} \textbf{\bibinfo{volume}{415}},
  \bibinfo{pages}{412} (\bibinfo{year}{2002}).

\bibitem[{\citenamefont{Kohsaka et~al.}(2004)\citenamefont{Kohsaka, Iwaya,
  Satow, Hanaguri, Azuma, Takano, and Takagi}}]{Kohsaka04}
\bibinfo{author}{\bibfnamefont{Y.}~\bibnamefont{Kohsaka}},
  \bibinfo{author}{\bibfnamefont{K.}~\bibnamefont{Iwaya}},
  \bibinfo{author}{\bibfnamefont{S.}~\bibnamefont{Satow}},
  \bibinfo{author}{\bibfnamefont{T.}~\bibnamefont{Hanaguri}},
  \bibinfo{author}{\bibfnamefont{M.}~\bibnamefont{Azuma}},
  \bibinfo{author}{\bibfnamefont{M.}~\bibnamefont{Takano}}, \bibnamefont{and}
  \bibinfo{author}{\bibfnamefont{H.}~\bibnamefont{Takagi}},
  \bibinfo{journal}{Phys. Rev. Lett.} \textbf{\bibinfo{volume}{93}},
  \bibinfo{pages}{097004} (\bibinfo{year}{2004}).

\bibitem[{\citenamefont{Machtoub et~al.}()\citenamefont{Machtoub, Keimer, and
  Yamada}}]{re:machtoub05}
\bibinfo{author}{\bibfnamefont{L.}~\bibnamefont{Machtoub}},
  \bibinfo{author}{\bibfnamefont{B.}~\bibnamefont{Keimer}}, \bibnamefont{and}
  \bibinfo{author}{\bibfnamefont{K.}~\bibnamefont{Yamada}},
  \bibinfo{note}{cond-mat/0502524 (2005)}.

\bibitem[{\citenamefont{Lake et~al.}(2002)\citenamefont{Lake, Ronnow,
  Christensen, Aeppli, Lefmann, McMorrow, Vorderwisch, Smeibidl, Mangkorntong,
  Sasagawa et~al.}}]{re:lake02}
\bibinfo{author}{\bibfnamefont{B.}~\bibnamefont{Lake}},
  \bibinfo{author}{\bibfnamefont{H.}~\bibnamefont{Ronnow}},
  \bibinfo{author}{\bibfnamefont{N.}~\bibnamefont{Christensen}},
  \bibinfo{author}{\bibfnamefont{G.}~\bibnamefont{Aeppli}},
  \bibinfo{author}{\bibfnamefont{K.}~\bibnamefont{Lefmann}},
  \bibinfo{author}{\bibfnamefont{D.}~\bibnamefont{McMorrow}},
  \bibinfo{author}{\bibfnamefont{P.}~\bibnamefont{Vorderwisch}},
  \bibinfo{author}{\bibfnamefont{P.}~\bibnamefont{Smeibidl}},
  \bibinfo{author}{\bibfnamefont{N.}~\bibnamefont{Mangkorntong}},
  \bibinfo{author}{\bibfnamefont{T.}~\bibnamefont{Sasagawa}},
  \bibnamefont{et~al.}, \bibinfo{journal}{Nature}
  \textbf{\bibinfo{volume}{415}}, \bibinfo{pages}{299} (\bibinfo{year}{2002}).

\bibitem[{\citenamefont{Dordevic et~al.}(2003)\citenamefont{Dordevic, Komiya,
  Ando, and Basov}}]{re:dordevic03}
\bibinfo{author}{\bibfnamefont{S.}~\bibnamefont{Dordevic}},
  \bibinfo{author}{\bibfnamefont{S.}~\bibnamefont{Komiya}},
  \bibinfo{author}{\bibfnamefont{Y.}~\bibnamefont{Ando}}, \bibnamefont{and}
  \bibinfo{author}{\bibfnamefont{D.}~\bibnamefont{Basov}},
  \bibinfo{journal}{Phys. Rev. Lett.} \textbf{\bibinfo{volume}{91}},
  \bibinfo{pages}{167401} (\bibinfo{year}{2003}).

\bibitem[{\citenamefont{Komiya and Ando}(2004)}]{re:komiya04}
\bibinfo{author}{\bibfnamefont{S.}~\bibnamefont{Komiya}} \bibnamefont{and}
  \bibinfo{author}{\bibfnamefont{Y.}~\bibnamefont{Ando}},
  \bibinfo{journal}{Phys. Rev. B} \textbf{\bibinfo{volume}{70}},
  \bibinfo{pages}{060503(R)} (\bibinfo{year}{2004}).

\bibitem[{\citenamefont{Keren et~al.}()\citenamefont{Keren, Kanigel, Lord, and
  Amato}}]{re:keren02}
\bibinfo{author}{\bibfnamefont{A.}~\bibnamefont{Keren}},
  \bibinfo{author}{\bibfnamefont{A.}~\bibnamefont{Kanigel}},
  \bibinfo{author}{\bibfnamefont{J.}~\bibnamefont{Lord}}, \bibnamefont{and}
  \bibinfo{author}{\bibfnamefont{A.}~\bibnamefont{Amato}},
  \bibinfo{note}{cond-mat/0211405 (2002)}.

\bibitem[{\citenamefont{Emery et~al.}(1990)\citenamefont{Emery, Kivelson, and
  Lin}}]{re:emery90}
\bibinfo{author}{\bibfnamefont{V.}~\bibnamefont{Emery}},
  \bibinfo{author}{\bibfnamefont{S.}~\bibnamefont{Kivelson}}, \bibnamefont{and}
  \bibinfo{author}{\bibfnamefont{H.}~\bibnamefont{Lin}},
  \bibinfo{journal}{Phys. Rev. Lett.} \textbf{\bibinfo{volume}{64}},
  \bibinfo{pages}{475} (\bibinfo{year}{1990}).

\bibitem[{\citenamefont{Poilblanc and Rice}(1989)}]{re:poilblanc89}
\bibinfo{author}{\bibfnamefont{D.}~\bibnamefont{Poilblanc}} \bibnamefont{and}
  \bibinfo{author}{\bibfnamefont{T.}~\bibnamefont{Rice}},
  \bibinfo{journal}{Phys. Rev. B} \textbf{\bibinfo{volume}{39}},
  \bibinfo{pages}{9749} (\bibinfo{year}{1989}).

\bibitem[{\citenamefont{Emery and Kivelson}(1993)}]{re:emery93}
\bibinfo{author}{\bibfnamefont{V.}~\bibnamefont{Emery}} \bibnamefont{and}
  \bibinfo{author}{\bibfnamefont{S.}~\bibnamefont{Kivelson}},
  \bibinfo{journal}{Physica C} \textbf{\bibinfo{volume}{209}},
  \bibinfo{pages}{597} (\bibinfo{year}{1993}).

\bibitem[{\citenamefont{Zaanen and Gunnarsson}(1989)}]{re:zaanen89}
\bibinfo{author}{\bibfnamefont{J.}~\bibnamefont{Zaanen}} \bibnamefont{and}
  \bibinfo{author}{\bibfnamefont{O.}~\bibnamefont{Gunnarsson}},
  \bibinfo{journal}{Phys. Rev. B} \textbf{\bibinfo{volume}{40}},
  \bibinfo{pages}{R7391} (\bibinfo{year}{1989}).

\bibitem[{\citenamefont{Emery and Kivelson}(1994)}]{re:emery94}
\bibinfo{author}{\bibfnamefont{V.}~\bibnamefont{Emery}} \bibnamefont{and}
  \bibinfo{author}{\bibfnamefont{S.}~\bibnamefont{Kivelson}},
  \bibinfo{journal}{Physica C} \textbf{\bibinfo{volume}{235}},
  \bibinfo{pages}{189} (\bibinfo{year}{1994}).

\bibitem[{\citenamefont{Tranquada et~al.}(1995)\citenamefont{Tranquada,
  Sternlieb, Axe, Nakamura, and Uchida}}]{re:tranquada95}
\bibinfo{author}{\bibfnamefont{J.~M.} \bibnamefont{Tranquada}},
  \bibinfo{author}{\bibfnamefont{B.~J.} \bibnamefont{Sternlieb}},
  \bibinfo{author}{\bibfnamefont{J.~D.} \bibnamefont{Axe}},
  \bibinfo{author}{\bibfnamefont{Y.}~\bibnamefont{Nakamura}}, \bibnamefont{and}
  \bibinfo{author}{\bibfnamefont{S.}~\bibnamefont{Uchida}},
  \bibinfo{journal}{Nature} \textbf{\bibinfo{volume}{375}},
  \bibinfo{pages}{561} (\bibinfo{year}{1995}).

\bibitem[{\citenamefont{Ghosal et~al.}(2002)\citenamefont{Ghosal, Kallin, and
  Berlinsky}}]{re:ghosal02}
\bibinfo{author}{\bibfnamefont{A.}~\bibnamefont{Ghosal}},
  \bibinfo{author}{\bibfnamefont{C.}~\bibnamefont{Kallin}}, \bibnamefont{and}
  \bibinfo{author}{\bibfnamefont{A.~J.} \bibnamefont{Berlinsky}},
  \bibinfo{journal}{Phys. Rev. B} \textbf{\bibinfo{volume}{66}},
  \bibinfo{pages}{214502} (\bibinfo{year}{2002}).

\bibitem[{\citenamefont{Atkinson et~al.}(2003)\citenamefont{Atkinson,
  Hirschfeld, and Zhu}}]{re:atkinson03}
\bibinfo{author}{\bibfnamefont{W.}~\bibnamefont{Atkinson}},
  \bibinfo{author}{\bibfnamefont{P.}~\bibnamefont{Hirschfeld}},
  \bibnamefont{and} \bibinfo{author}{\bibfnamefont{L.}~\bibnamefont{Zhu}},
  \bibinfo{journal}{Phys. Rev. B} \textbf{\bibinfo{volume}{68}},
  \bibinfo{pages}{054501} (\bibinfo{year}{2003}).

\bibitem[{\citenamefont{Ichioka and Machida}(1999)}]{re:ichioka99}
\bibinfo{author}{\bibfnamefont{M.}~\bibnamefont{Ichioka}} \bibnamefont{and}
  \bibinfo{author}{\bibfnamefont{K.}~\bibnamefont{Machida}},
  \bibinfo{journal}{J. Phys. Soc. Jpn.} \textbf{\bibinfo{volume}{68}},
  \bibinfo{pages}{4020} (\bibinfo{year}{1999}).

\bibitem[{\citenamefont{Moraghebi et~al.}(2002)\citenamefont{Moraghebi, Yunoki,
  and Moreo}}]{re:moraghebi02}
\bibinfo{author}{\bibfnamefont{M.}~\bibnamefont{Moraghebi}},
  \bibinfo{author}{\bibfnamefont{S.}~\bibnamefont{Yunoki}}, \bibnamefont{and}
  \bibinfo{author}{\bibfnamefont{A.}~\bibnamefont{Moreo}},
  \bibinfo{journal}{Phys. Rev. Lett.} \textbf{\bibinfo{volume}{88}},
  \bibinfo{pages}{187001} (\bibinfo{year}{2002}).

\bibitem[{\citenamefont{M.~Moraghebi and Moreo}(2001)}]{re:moraghebi01}
\bibinfo{author}{\bibfnamefont{S.~Y.} \bibnamefont{M.~Moraghebi},
  \bibfnamefont{C.~Buhler}} \bibnamefont{and}
  \bibinfo{author}{\bibfnamefont{A.}~\bibnamefont{Moreo}},
  \bibinfo{journal}{Phys. Rev. B} \textbf{\bibinfo{volume}{63}},
  \bibinfo{pages}{214513} (\bibinfo{year}{2001}).

\bibitem[{\citenamefont{Vershinin et~al.}(2004)\citenamefont{Vershinin, Misra,
  Ono, Abe, Ando, and Yazdani}}]{re:vershinin04}
\bibinfo{author}{\bibfnamefont{M.}~\bibnamefont{Vershinin}},
  \bibinfo{author}{\bibfnamefont{S.}~\bibnamefont{Misra}},
  \bibinfo{author}{\bibfnamefont{S.}~\bibnamefont{Ono}},
  \bibinfo{author}{\bibfnamefont{Y.}~\bibnamefont{Abe}},
  \bibinfo{author}{\bibfnamefont{Y.}~\bibnamefont{Ando}}, \bibnamefont{and}
  \bibinfo{author}{\bibfnamefont{A.}~\bibnamefont{Yazdani}},
  \bibinfo{journal}{Science} \textbf{\bibinfo{volume}{303}},
  \bibinfo{pages}{1995} (\bibinfo{year}{2004}).

\end{thebibliography}

\end{document}